\documentclass[12pt]{article}

\usepackage{graphicx,psfrag,epsf,color}
\usepackage{amsmath,amssymb,amsfonts}
\usepackage{array}
\usepackage{cite}
\usepackage{multirow}
\usepackage{rotating}
\usepackage{slashed}
\usepackage{bm}
\usepackage{graphbox}

\usepackage{enumerate}
\usepackage{braket}

\usepackage{tensor}
\usepackage{mathtools}
\usepackage[colorlinks]{hyperref}
\usepackage{slashed}
\usepackage[dvipsnames]{xcolor}
\usepackage[capitalise]{cleveref}
\usepackage{dsfont}
\usepackage{simpler-wick}
\hypersetup{pageanchor=false,linkcolor=NavyBlue,citecolor=NavyBlue,urlcolor=RoyalPurple}

\definecolor{navy}{rgb}{0.0,0.0,0.5}
\definecolor{babyblue}{rgb}{0.5, 0.74, 0.99}
\definecolor{coolblack}{rgb}{0.0, 0.18, 0.39}

\newcounter{MBQ}

\newcounter{PHQ}

\newcounter{DSQ}

\newcounter{PHE}

\newcounter{DSE}

\newcommand{\be}{\begin{equation}}
\newcommand{\ee}{\end{equation}}
\newcommand{\bea}{\begin{eqnarray}}
\newcommand{\eea}{\end{eqnarray}}
\newcommand{\bi}{\begin{itemize}}
\newcommand{\ei}{\end{itemize}}
\newcommand{\ben}{\begin{enumerate}}
\newcommand{\een}{\end{enumerate}}
\newcommand{\bt}{\begin{tabular}}
\newcommand{\et}{\end{tabular}}
\newcommand{\lc}{\left[}
\newcommand{\rc}{\right]}
\newcommand{\lp}{\left(}
\newcommand{\rp}{\right)}

\newcommand{\np}{n_+}
\newcommand{\nm}{n_-}

\def\de{\mathrm{d}}

\newcommand{\nip}{n_{i+}}
\newcommand{\nim}{n_{i-}}
\definecolor{navy}{rgb}{0.0,0.0,0.5}

\newcommand{\nn}{\nonumber}

\usepackage{scalerel,stackengine,amsmath}

\newcommand{\innfwhat}[2]{%
  \styletofont{#1}%
  \dimen0 \fontcharic\next1 \skewchar\next1
  \advance\dimen0 -\fontcharic\next1`#2%
  \makebox[0pt][l]{$#1#2$}%
  \makebox[\widthof{$#1#2$}]{$#1\kern.5\dimen0 \widehat{\vphantom{#2}}$}}

\begin{document}
\allowdisplaybreaks

\begin{titlepage}

\begin{flushright}
{\small
TUM-HEP-1434/22\\ 
MITP-22-097\\
March 09, 2023 \\
arXiv:2212.02525 [hep-ph]
}
\end{flushright}

\vskip1cm
\begin{center}
{\Large \bf Soft-collinear gravity with fermionic matter}
\end{center}
  \vspace{0.5cm}
\begin{center}
{\sc Martin~Beneke,$^{a}$ \sc Patrick~Hager,$^{b}$ and Dominik~Schwienbacher$^{a}$} 
\\[6mm]
{\it ${}^a$Physik Department T31,\\
James-Franck-Stra\ss e~1,
Technische Universit\"at M\"unchen,\\
D--85748 Garching, Germany}\\[0.2cm]
{\it ${}^b$PRISMA\textsuperscript{+} Cluster of Excellence \& Mainz Institute for Theoretical Physics\\
Johannes Gutenberg University, D--55099 Mainz, Germany }
\end{center}
\vskip1cm

\begin{abstract}
\noindent 
We extend the effective field theory for soft and collinear 
gravitons to interactions with fermionic matter fields.
The full theory features a local Lorentz symmetry in addition to the usual diffeomorphisms, which requires incorporating the former into the soft-collinear gravity framework. The local Lorentz symmetry gives rise to Wilson lines in the effective theory that strongly resemble those in SCET for non-abelian gauge interactions, whereas the diffeomorphisms can be treated in the same fashion as in the case of scalar matter. The basic structure of soft-collinear gravity, which features a homogeneous soft background field, giving rise to a covariant derivative and multipole-expanded covariant Riemann-tensor interactions, remains unaltered and generalises in a natural way to fermion fields.
\end{abstract}

\end{titlepage}


\section{Introduction}

The soft and collinear limits of quantum field theories often
provide valuable structural insights, in particular for gauge theories
and gravitation. The soft (or eikonal) limit of
gauge theory and gravity is indeed very similar \cite{Weinberg:1965nx},
with gauge charge replaced by the gravitational
charge, momentum. The collinear limit is, however, very
different. While jets are a distinctive feature of gauge theories,
collinear divergences are absent in gravity
\cite{Weinberg:1965nx, Akhoury:2011kq, Beneke:2012xa},
since the angular pattern for the emission of massless
spin-2 particles suppresses forward emission.

The study of collinear graviton interactions is therefore
intrinsically a next-to-leading power problem.
Soft-collinear gravity \cite{Beneke:2012xa} is the effective theory
encompassing the interactions of soft and collinear gravitons with matter
particles and among themselves, in
principle to any power in the soft or collinear limit.
The general framework for scalar (and integer-spin tensorial)
representations, based on the underlying
diffeomorphism-invariance,\footnote{Also called ``general coordinate
transformations'' (GCT) in this context.} is derived systematically beyond
leading order in the power-counting parameter $\lambda$ in
\cite{Beneke:2021aip}.
It is based on the position-space formulation
\cite{Beneke:2002ph,Beneke:2002ni} of soft-collinear effective theory
(SCET) originally developed for QCD \cite{Bauer:2000yr,Bauer:2001yt}.
Here, we extend the construction to incorporate fields of half-integer
spin, presenting the derivation explicitly to second order in $\lambda$
for  Dirac fermions. Since half-integer spin fields require special
treatment in general relativity, the present work completes the framework
developed in \cite{Beneke:2021aip}.

For scalar and integer-spin fields, the gravitational interactions are diffeomorphism-invariant, and the respective redefinitions are well-understood to all orders in $\lambda$ \cite{Beneke:2021aip}.
For half-integer spin fields, however, the gravitational theory features an additional local Lorentz symmetry implemented by the vierbein formalism, for an introduction see e.g. \cite{weinberg_gravitation_book_vierbein,Misner:1973prb}. 
Consequently, the key ingredient to incorporate half-integer representations is the proper treatment of the local Lorentz symmetry. 
Unlike in the case of diffeomorphisms, the local Lorentz symmetry takes the form of a standard gauge symmetry. Our approach will be to treat fermions as scalars under diffeomorphisms, but implement local Lorentz transformations into SCET in analogy with gauge transformations in QCD with fermions transforming in a non-trivial representation of the local Lorentz symmetry.

The analogue of the non-abelian matrix-valued gauge field $A_\mu$ is the spin-connection $\Omega_\mu$.
Accordingly, one can define a Wilson line
\begin{equation}\label{eq:StandardLLTWilsonLine}
    W(z,x)\equiv \textbf{P}\exp\left[i\int_{\mathcal{C}(x,z)}  \de y^\mu\: \Omega_\mu(y)\right]\,,
\end{equation}
where $\textbf{P}$ denotes path-ordering along a curve $\mathcal{C}$ from $x$ to $z$, in complete analogy to gauge theory.
Under the local Lorentz symmetry, the Wilson line transforms as
\begin{equation}
\label{eq:LLT_trafo_Wilson}
    W(z,x)\rightarrow D_{\Lambda}(z)W(z,x)D^{-1}_{\Lambda}(x) \,,
\end{equation}
which precisely corresponds to the transformation of a gauge-field Wilson line.
Here, $D_\Lambda(x)$ is a local Lorentz transformation in a given representation, e.g.~spinorial $[D_\Lambda(x)]_{\alpha\beta}$ or vectorial $[D_\Lambda(x)]\indices{^a_b} \equiv \Lambda\indices{^a_b}(x)$, depending on the representation of $\Omega_\mu(x)$.\footnote{In the following, we suppress the spinor indices $\alpha,\beta,\dots$ to shorten the notation.}
These Wilson lines lie at the heart of the systematic construction of SCET to subleading order \cite{Beneke:2002ni}, since they can be used to construct composite operators that satisfy certain gauge-constraints without restricting the gauge-freedom in the Lagrangian.
In the following, the discussion focuses on the local Lorentz transformations (LLT), giving only a brief review of the treatment of diffeomorphisms to make the presentation self-contained. For further details on these, we refer to \cite{Beneke:2021aip,Beneke:2022pue}.

The massless Dirac fermion in curved space-time is described by the action\footnote{We employ the sign convention $(+,-,-,-)$ and work in units where the gravitational coupling is $\kappa=1$. Latin indices $a,b,\dots$ denote local Lorentz indices, while Greek ones denote general coordinate indices, indicating transformation under diffeomorphisms. The Latin indices are raised using the flat metric $\eta\indices{^a^b}$ while Greek indices are raised with the curved-space metric $g^{\mu\nu}(x)$.}
\begin{equation}\label{eq:DiracAction}
    S_m = \int \de^4x\sqrt{-g(x)}\; \overline{\psi}(x) \tensor{E}{^\mu_a}(x)\gamma^a i D_\mu\psi(x)\,,
\end{equation}
where $g(x)$ is the metric determinant, $\tensor{E}{^\mu_a}(x)$ the inverse vierbein and 
\begin{equation}
    D_\mu = \partial_\mu - i\Omega_\mu(x)
\end{equation}
the local Lorentz-covariant derivative.
The vierbein and its inverse are defined by 
\begin{equation}\label{eq:VierbeinDefinition}
    g_{\mu\nu}(x) = \tensor{e}{^a_\mu}(x)\tensor{e}{^b_\nu}(x)\,\eta_{ab}\,,\quad \tensor{E}{^\mu_a}(x)\tensor{e}{^a_\nu}(x) = \delta\indices{^\mu_\nu}\,,
\end{equation}
and the spin-connection is expressed in terms of the vierbein as
\begin{equation}
\label{eq:defining_spin_connection}
    \Omega_\mu(x) = \frac 12 \Sigma^{ab} \,\omega_{\mu ab}(x) = 
\frac 12 \Sigma^{ab} g_{\nu\rho}(x)\tensor{E}{^\rho_{a}}(x)\lc \nabla_\mu \tensor{E}{^\nu_b}(x)\rc\,,
\end{equation}
where $\Sigma^{ab}$ is the generator of 
the representation of the Lorentz group, in which the matrix-valued spin connection $\Omega_\mu(x)$ is evaluated. Here $\Sigma^{ab} = \frac{i}{4}[\gamma^a,\gamma^b]$ for the spin-$\frac{1}{2}$ Dirac representation. $\nabla_\mu$ is the GCT-covariant derivative given by
\begin{equation}
    \nabla_\mu \tensor{E}{^\nu_b}(x) = \partial_\mu \tensor{E}{^\nu_b}(x) + \tensor{\Gamma}{^\nu_{\mu\alpha}}(x)\tensor{E}{^\alpha_b}(x)\,. 
\end{equation}
In order to quantise the metric fluctuations, the metric tensor is weak-field expanded about flat-space with Minkowski metric $\eta_{\mu\nu}$ as
\begin{equation}
    g_{\mu\nu}(x) = \eta_{\mu\nu} + h_{\mu\nu}(x)\,.
\end{equation}
Due to \eqref{eq:VierbeinDefinition}, this also induces the weak-field expansion of the vierbein and its inverse.
However, this condition does not uniquely determine the vierbein, since the defining equation in \eqref{eq:VierbeinDefinition} is still invariant under a LLT
\begin{equation}
    \tensor{e}{^a_\mu}(x)\tensor{e}{^b_\nu}(x)\,\eta_{ab} \to \tensor{e}{^a_\mu}(x)\tensor{e}{^b_\nu}(x)\tensor{\Lambda}{_a^c}(x)\tensor{\Lambda}{_b^d}(x)\,\eta_{cd} = \tensor{e}{^a_\mu}(x)\tensor{e}{^b_\nu}(x)\,\eta_{ab}\,.
\end{equation}
Therefore, it is convenient to factorise the local Lorentz contribution from the vierbein as
\begin{equation}
\label{eq:expandedVierbein}
    e\indices{^a_\mu}(x)= \Xi\indices{^a_b}(x) [e_{\rm sym}]\indices{^b_\mu}(x)\,,
\end{equation}
where $[e_{\rm sym}]\indices{^b_\mu}(x)$ is the ``symmetric vierbein'', defined by its weak-field expansion in terms of $h_{\mu\nu}(x)$ as
\begin{equation}\label{eq:hate}
        [e_{\rm sym}]\indices{^b_\mu}(x)=\delta\indices{^b_\mu}+\frac{1}{2}h\indices{^b_\mu}(x) -\frac{1}{8} h\indices{^b_\alpha}(x)h\indices{^\alpha_\mu}(x) + \mathcal{O}(h^3)\,.
\end{equation}
In this decomposition \eqref{eq:expandedVierbein}, the symmetric vierbein is uniquely determined by the metric fluctuation $h_{\mu\nu}(x)$ through \eqref{eq:hate}.
Moreover, the symmetric vierbein $[e_{\rm sym}]\indices{^b_\mu}(x)$ is LLT-invariant, and
the action of the local Lorentz group is consequently allocated entirely to $\tensor{\Xi}{^a_b}(x)$.
This arises because the metric fluctuation $h_{\mu\nu}(x)$ contains 10 dynamic degrees of freedom, which constrain only 10 of the 16 components of the vierbein $e\indices{^a_\mu}(x)$. 
The remaining 6 components parameterise the local Lorentz symmetry and constitute $\Xi\indices{^a_b}(x)$. These ``antisymmetric" components of the vierbein are not dynamic, as the Einstein-Hilbert action may be expressed purely in terms of the metric field itself.
Fixing the LLT gauge therefore reduces to determining the matrix $\Xi\indices{^a_b}(x)$ in \eqref{eq:expandedVierbein}.
The most convenient gauge condition is ``symmetric gauge'', where one simply chooses $\Xi\indices{^a_b}(x) = \delta\indices{^a_b}$.
Note that for the symmetric vierbein \eqref{eq:hate} in its weak-field expanded form, we no longer distinguish LLT and GCT indices, as all indices are now raised and lowered with the flat-space metric $\eta_{\mu\nu}$.
At this point, one might be concerned  that information is lost or the vierbein has the wrong transformation, however, this is not the case.
Infinitesimally, the GCT transformation $h_{\mu\nu}(x) \to h_{\mu\nu}(x) - \partial_\mu\varepsilon_\nu(x) - \partial_\nu \varepsilon_\mu(x)$ implies
\begin{equation}
    \delta\indices{^a_\mu}+\frac{1}{2}h\indices{^a_\mu}(x) \to \delta\indices{^a_\mu}+\frac{1}{2}h\indices{^a_\mu}(x) - \frac 12\partial_\mu\varepsilon^a(x) - \frac 12 \partial^a\varepsilon_\mu(x) + \mathcal{O}(h^2,\varepsilon^2,h\varepsilon)\,.
\end{equation}
This is \emph{not} the infinitesimal form of a covector transformation. However, by rewriting
\begin{equation}
    -\frac 12\partial_\mu\varepsilon^a(x) - \frac 12\partial^a\varepsilon_\mu(x) = -\partial_\mu\varepsilon^a(x) - \frac 12\bigl(\partial^a\varepsilon_\mu(x) - \partial_\mu \varepsilon^a(x)\bigr) \equiv -\partial_\mu\varepsilon^a(x) - \lambda\indices{^a_\mu}(x)\,, 
\end{equation}
one finds that the diffeomorphism transformation also generates a local Lorentz transformation with parameter $\lambda\indices{^a_\mu}(x)$ that depends on the derivative of the GCT gauge parameter $\varepsilon^\mu(x)$.
The vierbein $[e_{\rm sym}]\indices{^a_\mu}(x)$ therefore has the infinitesimal transformation
\begin{equation}
    [e_{\rm sym}]\indices{^a_\mu}(x) \to [e_{\rm sym}]\indices{^a_\mu}(x) - \partial_\mu\varepsilon^\alpha(x) [e_{\rm sym}]\indices{^a_\alpha}(x) - \lambda\indices{^a_b}(x)[e_{\rm sym}]\indices{^b_\mu}(x) + \mathcal{O}(\varepsilon^2)\,,
\end{equation}
and gauge-invariance is guaranteed by the diffeomorphism invariance and local Lorentz invariance of the full-theory expressions.
This also indicates that in the LLT gauge-fixed theory, one can still treat the Latin indices as LLT ones, but there is no distinction between LLT and GCT indices anymore, as the only residual symmetry left in the weak-field expansion are the diffeomorphisms---the diffeomorphisms also generate the local Lorentz transformation. 

The choice of the LLT gauge-condition has subtle consequences for scattering amplitudes and is relevant when comparing results obtained in different gauges, e.g. when performing a matching computation.
Since the gauge-fixing is equivalent to choosing a certain local inertial frame, the condition affects the external polarisation states that enter the definition of the scattering amplitude.
To avoid these subtleties altogether, it is convenient to directly fix symmetric gauge in both the full theory and the effective theory, which we do in Sec.~\ref{sec:EmergentBG} once the Wilson lines are obtained.

\section{SCET construction}

To construct SCET, one splits the full-theory fields into a soft mode and (multiple) collinear modes.
Each collinear sector comes with its respective light-cone vectors $n_{i\pm}^\mu$ satisfying $\nip^2=\nim^2=0$ and $\nip\cdot\nim=2$. With respect to these, the components of collinear momenta scale as $(\nip p,p_\perp,\nim p)\sim(1,\lambda,\lambda^2).$\footnote{Since only a single collinear direction appears in the Lagrangian, we drop the subscript $i$ in the following.}
Soft momenta scale isotropically as $k\sim\lambda^2.$
For spinorial matter fields, one additionally defines the projectors
\begin{equation}
\label{eq:projection_operators}
    P_+ = \frac{\slashed{n}_-\slashed{n}_+}{4}\,,\quad P_- = \frac{\slashed{n}_+\slashed{n}_-}{4}\,,
\end{equation}
and decomposes the collinear spinor-field $\psi_c$ as
\begin{equation}\label{eq:SpinorSplit}
    \psi_c(x) = \frac{\slashed{n}_-\slashed{n}_+}{4}\psi_c(x) + \frac{\slashed{n}_+\slashed{n}_-}{4}\psi_c(x) \equiv \xi(x) + \zeta(x)\,. 
\end{equation}
These components then scale as $\xi\sim\lambda,$ $\zeta\sim\lambda^2$ \cite{Beneke:2002ni}. In addition, the theory contains a soft fermion field $q\sim\lambda^3$.
For the metric tensor, it is convenient to consider the split \cite{Beneke:2021aip}
\begin{equation}
\label{eq:metric_split}
    g_{\mu\nu}(x) = g_{s\mu\nu}(x) + h_{\mu\nu}(x)\,,
\end{equation}
into a collinear fluctuation $h_{\mu\nu}(x)$ about a soft background $g_{s\mu\nu}(x)\equiv \eta_{\mu\nu} + s_{\mu\nu}(x)$.
The power-counting of the collinear graviton field is given by $h_{\mu\nu}\sim\frac{p_\mu p_\nu}{\lambda}$, while the soft graviton field scales as $s\indices{_\mu_\nu}\sim\lambda^2$ for all components \cite{Beneke:2012xa}.
Accordingly, it is convenient to perform a similar split also for the vierbein
\begin{equation}\label{eq:vierbeinsplit}
    e\indices{^a_\mu}(x)=e\indices{_s^a_\mu}(x)+e\indices{_c^a_\mu}(x)
\end{equation}
where the soft vierbein $e\indices{_s^a_\mu}(x)$ satisfies
\begin{equation}
\label{eq:vierbein_definition}
    g_{s\mu\nu}(x)=e\indices{_s^a_\mu}(x)e\indices{_s^b_\nu}(x)\eta_{ab}\,,
\end{equation}
and for the spin-connection
\begin{equation}\label{eq:SpinConnectionSplit}
    \Omega_\mu(x) = \Omega_{s\mu}(x) + \Omega_{c\mu}(x)\,.
\end{equation}
Note that while $h_{\mu\nu}(x)$ is defined in \eqref{eq:metric_split} to be a purely-collinear object, the collinear vierbein fluctuation\footnote{Note that while $e\indices{_s^a_\mu}(x)$ is the vierbein analogue of $g\indices{_s_\mu_\nu}(x)$ and not $s\indices{_\mu_\nu}(x)$, $e\indices{_c^a_\mu}(x)$ corresponds to the collinear metric {\em fluctuation} $h\indices{_\mu_\nu}(x)$.} $e\indices{_c^a_\mu}(x)$ determined from \eqref{eq:VierbeinDefinition}, \eqref{eq:metric_split}, 
\eqref{eq:vierbein_definition} as well as the collinear spin-connection $\Omega_{c\mu}(x)$ in \eqref{eq:SpinConnectionSplit} are not purely-collinear objects beyond leading-power, since the soft background metric appears in their definition.\footnote{This is similar to the inverse metric tensor
    $g^{\mu\nu} = g_s^{\mu\nu} - g_s^{\mu\alpha} g_s^{\nu\beta} h_{\alpha\beta} + \mathcal{O}(h^2)$,
where the ``inverse fluctuation'' $g_s^{\mu\alpha}g_s^{\nu\beta} h_{\alpha\beta} + \dots$ is also not a purely-collinear object. }
However, the soft counterparts are always purely-soft objects.
Since both the vierbein and the spin-connection only serve as auxiliary quantities, this is not a problem for the following discussion.
The physical graviton polarisations can be clearly separated into collinear and soft modes.

At this point, one could employ symmetric gauge to uniquely determine the vierbein. However, to systematically construct soft-collinear gravity, one exploits the local Lorentz symmetry by fixing different gauges in the intermediate steps and reintroduces the full gauge symmetry using Wilson lines \cite{Beneke:2021aip, Beneke:2002ni}. Only once this construction is complete and the redefinitions are performed, one fixes symmetric gauge in the full and the effective theory to facilitate the matching computations.

In the position-space formalism, the coordinate arguments of the fields scale as $n_{-} x \sim 1,$ $x_\perp\sim\frac{1}{\lambda},$ $\np x \sim \frac{1}{\lambda^2}$ for collinear and $x\sim\frac{1}{\lambda^2}$ for soft fields.
To obtain a homogeneous Lagrangian leading to a systematic soft-collinear expansion, the different scaling of collinear and soft coordinates necessitates a multipole expansion in soft-collinear interactions about $x_-^\mu = \np x \frac{\nm^\mu}{2}$ \cite{Beneke:2002ph}, which generates an infinite tower of subleading terms.
This is the position-space analogue of expanding the scattering amplitude in the small soft momenta.
For the systematic construction of the effective theory,
the essential step is to redefine the collinear fields such that their soft gauge transformation respects this multipole expansion.
This cements the underlying gauge symmetry as the foundation on which the effective theory is constructed.

As a consequence of the soft and collinear mode split, the GCT and LLT themselves split into soft and collinear transformations.
Soft fields cannot transform under collinear gauge transformations, otherwise they would acquire collinear fluctuations and no longer be soft.
However, the right-hand sides of the splits \eqref{eq:metric_split}, \eqref{eq:vierbeinsplit} must transform as the full-theory fields on the left-hand sides.
This constrains the form of the gauge transformations.
The precise transformations under collinear and soft diffeomorphisms read\footnote{For GCT, both $\xi$ and $\zeta$ transform like a scalar field.} \cite{Beneke:2021aip}
    \begin{align}
      &\text{collinear:} &  h\indices{_\mu_\nu}&\rightarrow\left[U_cU\indices{_c_\mu^\alpha}U\indices{_c_\nu^\beta}\left(g\indices{_s_\alpha_\beta}+h\indices{_\alpha_\beta}\right)\right]-g\indices{_s_\mu_\nu}\,, & 
      \psi_c&\rightarrow\left[U_c\psi_c\right]\,,\nn\\
       & & e\indices{_c^a_\mu}&\rightarrow \left[U_cU\indices{_c_\mu^\alpha}\left(e\indices{_s^a_\alpha}+e\indices{_c^a_\alpha}\right)\right]-e\indices{_s^a_\mu}\,,\label{eq:GCTcoll}\\
       & & s\indices{_\mu_\nu}&\rightarrow s\indices{_\mu_\nu}\,, & q &\rightarrow q\nn\\
       & & e\indices{_s^a_\mu}&\rightarrow e\indices{_s^a_\mu}\,,\nn\\
       &\text{soft:} & h\indices{_\mu_\nu}(x)&\rightarrow\left[U_s(x)U\indices{_s_\mu^\alpha}(x)U\indices{_s_\nu^\beta}(x)h\indices{_\alpha_\beta}(x)\right]\,, &
        \psi_c(x)&\rightarrow \left[U_s(x)\psi_c(x)\right]\,,\nn\\
        & & e\indices{_c^a_\mu}(x)&\rightarrow \left[U_s(x)U\indices{_s_\mu^\alpha}(x) e\indices{_c^a_\alpha}(x)\right]\,,\label{eq:GCTsoft}\\
        & & s\indices{_\mu_\nu}&\rightarrow \left[U_sU\indices{_s_\mu^\alpha}U\indices{_s_\nu^\beta}(\eta_{\alpha\beta} + s_{\alpha\beta})\right] - \eta_{\mu\nu}\,, &
        q&\rightarrow \left[U_sq\right]\,,\nn\\
        & & e\indices{_s^a_\mu}&\rightarrow \left[U_sU\indices{_s_\mu^\alpha} e\indices{_s^a_\alpha}\right]\,.\nn
    \end{align}
    Here, we adopt the active point of view \cite{Beneke:2021aip}, whereby only the fields and not the coordinates transform under diffeomorphisms.
    We will adopt the same point of view for the local Lorentz transformations in the following.
    In particular, this implies that the reference vectors $n_{\pm}^\mu$ defined previously neither transform under GCT nor LLT but are fixed.\footnote{In particular, this implies that an object $\np^\mu A_\mu$ is \emph{not} GCT-invariant, but transforms like the $\nm^\mu$-component of a vector.
    In an abuse of notation, this also means that $\np^a = \np^\mu \delta^a_\mu$.}
    All fields and transformations in \eqref{eq:GCTcoll}, \eqref{eq:GCTsoft} are understood to be evaluated at $x$, but we have emphasised that the soft transformations of collinear fields still depend on the full collinear argument $x$, and consequently do not respect the multipole expansion by mixing different orders in $\lambda$.
    One can verify that the full-theory metric tensor $g_{s\mu\nu}+h_{\mu\nu}$ as well as the vierbein $\tensor{e}{_s^a_\mu}+\tensor{e}{_c^a_\mu}$ indeed have the standard diffeomorphism transformation of a metric tensor and vierbein (a covector), respectively.\footnote{The naive decomposition $\psi = \psi_c + q$ does not have the correct collinear gauge transformation of the full spinor on the left-hand side. The proper redefinition requires Wilson lines and is given in \eqref{eq:softquark} below.} 
    Note that the collinear transformation \eqref{eq:GCTcoll} of the collinear graviton field is defined around the soft background-metric $g_{s\mu\nu}(x)$, which appears in the transformation in place of $\eta_{\mu\nu}$. 
    
Observe that the soft metric fluctuation $s_{\mu\nu}$ has the standard transformation of a metric perturbation under soft transformations \eqref{eq:GCTsoft}. On the other hand, the collinear fluctuation $h_{\mu\nu}$ does not transform like a perturbation, but like an ordinary tensor field.
    
For local Lorentz transformations, the splits \eqref{eq:vierbeinsplit}, \eqref{eq:SpinConnectionSplit} yield the transformations\footnote{It is convenient to work with the full collinear field $\psi_c=\xi+\zeta$ \eqref{eq:SpinorSplit}, as only this sum has the standard transformation. The components $\xi$, $\zeta$ do not have a simple LLT, since the projectors \eqref{eq:projection_operators} do not commute with the LLT.}
\begin{align}
 &\text{collinear:} & 
      \Omega\indices{_c_\mu}&\to D_{\Lambda_c}\Omega\indices{_c_\mu}D_{\Lambda_c}^{-1} +iD_{\Lambda_c}\left[\partial_\mu-i\Omega\indices{_s_\mu}, D_{\Lambda_c}^{-1}\right]\,,\; &
 \psi_c &\rightarrow D_{\Lambda_c}\psi_c\,,\nn\\
 & &  \Omega\indices{_s_\mu}&\to\Omega\indices{_s_\mu}\,,\; &
q &\to q \label{eq:defining_collinear_LLT_trafos}\,, \\
& & e\indices{_c^a_\mu}&\to \Lambda\indices{_c^a_b}\left(e\indices{_s^b_\mu}+e\indices{_c^b_\mu}\right)-e\indices{_s^a_\mu}\,,\nn\\
& & e\indices{_s^a_\mu}&\to e\indices{_s^a_\mu}\,,\nn\\
&\text{soft:} &     
\Omega\indices{_c_\mu}(x)&\to  D_{\Lambda_s}(x)\Omega\indices{_c_\mu}(x) D_{\Lambda_s}^{-1}(x)\,,\; &
\psi_c(x)&\to D_{\Lambda_s}(x)\psi_c(x)\,,\nn\\
& & \Omega\indices{_s_\mu}&\to D_{\Lambda_s}\Omega\indices{_s_\mu}D_{\Lambda_s}^{-1}+iD_{\Lambda_s}\left[\partial_\mu D_{\Lambda_s}^{-1}\right]\,,\; 
& q&\to D_{\Lambda_s}q \label{eq:starting_soft_LLT_trafos}\,,\\
& & e\indices{_c^a_\mu}(x)&\to \Lambda\indices{_s^a_b}(x)e\indices{_c^b_\mu}(x)\,,\nn\\
& &  e\indices{_s^a_\mu}&\to \Lambda\indices{_s^a_b} \, e\indices{_s^b_\mu}\,.\nn
    \end{align}
Whereas the GCT \eqref{eq:GCTcoll}, \eqref{eq:GCTsoft} take a rather unconventional form, the LLT correspond to standard gauge transformations, see Eq.~(1) in \cite{Beneke:2002ni}.
In the collinear transformations of $\Omega_{c\mu}$, the last bracket can be rewritten using $\partial_\mu - i\Omega_{s\mu} = D_{s,\mu}$ as 
\begin{equation}
    \Omega_{c\mu} \to D_{\Lambda_c} \Omega_{c\mu} D_{\Lambda_c}^{-1} + i D_{\Lambda_s}\lc D_{s,\mu}, D_{\Lambda_c}^{-1}\rc\,,
\end{equation}
which is the exact analogue of the gluon transformation (1) in \cite{Beneke:2002ni}.
The treatment of LLT is therefore linked more closely to non-abelian gauge transformations than to the diffeomorphisms.
 
\section{``Homogeneous'' gauge transformations}

In interaction terms where soft and collinear fields are present, one must perform the light-front multipole expansion of soft fields $\phi_s(x)$ \cite{Beneke:2002ni}.
To obtain a Lagrangian containing homogeneous, gauge-covariant terms, the soft gauge transformation should be compatible with this expansion.
However, the soft transformations of collinear fields given in \eqref{eq:GCTsoft}, \eqref{eq:starting_soft_LLT_trafos} depend on the full collinear coordinate argument $x$, and do not respect the multipole expansion about $x_-$.
Instead, collinear fields should transform according to a ``homogeneous'' soft gauge symmetry that does not mix different orders in $\lambda$ due to the coordinate arguments of the parameters.
This is accomplished \cite{Beneke:2002ni} by a field redefinition of the collinear fields 
 $\hat{\psi}_c$ and $\hat{e}\indices{_c^a_\mu}$, such that the soft gauge transformation of the new fields respects the multipole-expansion for \emph{both} the diffeomorphism and the local Lorentz symmetry.
For LLT in particular, the fields evaluated at $x$ must transform as
\begin{equation}
\begin{aligned}
\label{eq:defining_soft_LLT_trafos}
        \hat{\Omega}\indices{_c_\mu}&\to  D_{\Lambda_s}(x_-)\hat{\Omega}\indices{_c_\mu} D_{\Lambda_s}^{-1}(x_-)\,,\; & \hat{\psi}_c&\to D_{\Lambda_s}(x_-)\hat{\psi}_c\,,\; &
         \hat{e}\indices{_c^a_\mu}&\to \Lambda\indices{_s^a_b}(x_-) \hat{e}\indices{_c^b_\mu}\,.
\end{aligned}
\end{equation}
To relate these redefined hatted fields to the original ones, one employs a number of Wilson lines that are introduced in the following sections.
However, before approaching this problem, one first has to decide in which order the redefinitions for diffeomorphisms and local Lorentz transformations should be performed, since a diffeomorphism corresponds to a translation $x\to x^\prime$, which affects the argument of the local Lorentz transformation. 
Therefore, either one first ensures that the diffeomorphisms respect the multipole expansion, and then constructs redefinitions that fix the LLT behaviour afterwards, or one does it in the opposite order.
Since the redefinitions for the diffeomorphisms are already worked out in \cite{Beneke:2021aip}, we employ these first and fix the LLT for the respective redefined fields.

In gauge theory \cite{Beneke:2002ni}, one employs two Wilson lines in the redefinition.
The first one is the collinear Wilson line $W_c$, defined as
\begin{equation}\label{eq:GaugeWC}
    W_c(x) = \textbf{P}\exp\lp ig_s\int_{-\infty}^0 \de s\: \np \hat{A}_c(x+sn_+)\rp\,.
\end{equation}
This Wilson line transforms as $W_c(x)\to U_c(x)W_c(x)$ and can be used to construct a collinear gauge-invariant building block, in particular the gauge-invariant spinor field $\hat{\chi}_c = W_c^\dagger\hat{\xi}_c$ or the gluon field $g_s\hat{\mathcal{A}}_c = W_c^\dagger i\hat{D}^\mu W_c - iD_s^\mu$.
This gauge-invariant field is a composite operator that fulfills the collinear light-cone gauge condition $\np \hat{\mathcal{A}}_c = 0$, and is necessary to control the large gluon components $\np A_c \sim 1$ as well as to 
construct the collinear field redefinition.

The second Wilson line is the soft $R(x)$, given by
\begin{equation}\label{eq:GaugeR}
    R(x) = \textbf{P}\exp\lp ig_s\int_{0}^1 \de s\: (x-x_-)^\mu A_{s\mu}\bigl(x+s(x-x_-)\bigr)\rp\,. 
\end{equation}
This Wilson line transports the gauge transformation from point $x_-$ to point $x$, as can be seen by its transformation
\begin{equation}
    R(x) \to U_s(x) R(x) U_s^\dagger(x_-)\,.
\end{equation}
It is used to relate the homogeneously transforming, hatted  fields to the original ones.
For example, for the collinear quark field, the field redefinition reads
\begin{equation}
    \xi_c(x) = R(x) W_c^\dagger(x)\hat{\xi}_c(x)\,,
\end{equation}
where the field on the left-hand side is taken to be in collinear light-cone gauge.
On the right-hand side, the collinear-gauge-invariant $\hat{\chi}_c(x) = W_c^\dagger(x)\hat{\xi}_c(x)$ appears and its soft gauge transformation is moved from $x_-$ to $x$ by virtue of $R(x)$, consistent with the transformation of the left-hand side.
This intuition and the two necessary Wilson lines carry through to GCT, as reviewed in the next paragraph, and also to the LLT, as presented in Sec.~\ref{sec:Wilson}.

For the GCT, the redefinitions relating the homogeneously GCT-transforming fields $\Tilde{e}, \Tilde{\psi}$ to the original ones are readily adapted from \cite{Beneke:2021aip}.\footnote{
In the following, we employ a different notation compared to \cite{Beneke:2021aip} for the fields: $\phi$ are the full-theory ones, $\tilde{\phi}$ correspond to the redefined collinear fields that transform homogeneously under GCT, denoted as $\hat{\phi}$ in \cite{Beneke:2021aip}.
We change the notation to $\tilde{\phi}$ to emphasise that we have not yet homogenised the additional LLT symmetry for fermions and the vierbein. The final fields, transforming homogeneously under both LLT and GCT, are denoted correspondingly as $\hat{\phi}$ in this article.}
In the place of the two Wilson lines in gauge theory \eqref{eq:GaugeWC}, \eqref{eq:GaugeR}, one employs the analogous objects for diffeomorphisms, which correspond to two local translation operators.
The first one is the collinear ``Wilson line'' $W_c^{-1}$, which takes the form
\begin{equation}
    W_c^{-1} = T_{\theta_{\rm LC}} = 1 + \theta_{\rm LC}^\alpha(x)\partial_\alpha + \frac 12 \theta_{\rm LC}^\alpha(x)\theta_{\rm LC}^\beta(x)\partial_\alpha\partial_\beta + \dots\,,
\end{equation}
as given in (5.101) in \cite{Beneke:2021aip}. 
This ``Wilson line'' can be viewed as the choice of a special 
coordinate system \cite{Donnelly:2015hta} or collinear diffeomorphism gauge-fixing. Its use 
here is to define a gauge-invariant composite 
graviton field $\mathfrak{h}_{\mu\nu}(x)$ that satisfies the collinear light-cone gauge property $\mathfrak{h}_{\mu+}(x)=0$ by the appropriate choice of $\theta_{\rm LC}^\alpha(x)$, which therefore depends on the given $h_{\mu\nu}(x)$. The gauge-invariant spinor-field then reads
\begin{equation}
    \left[W_c^{-1}\tilde{\psi}_c(x)\right] = \tilde{\psi}_c(x+\theta_{\rm LC}(x))\,.
\end{equation}
Since the vierbein transforms like a covariant vector under GCT, one can define the collinear-invariant vierbein similarly as 
\begin{equation}\label{eq:CollInvrVierbein}
   \tilde\epsilon\indices{_c^a_\mu}(x)=\left[W\indices{^\rho_\mu}W\indices{_c^{-1}}\left(\hat{e}\indices{_s^a_\rho}(x)+\tilde{e}\indices{_c^a_\rho}(x)\right) \right]-\hat{e}\indices{_s^a_\mu}(x) = \tilde{e}\indices{^a_\mu}(x+\theta_{\rm LC}(x)) - \hat{e}\indices{_s^a_\mu}(x)\,,
\end{equation}
where $\hat{e}\indices{_s^a_\mu}(x)$ is the vierbein defined in \eqref{eq:VierbeinSplitBG} below and derived in the corresponding section. In Sec.~\ref{sec:EmergentBG} it is shown that this vierbein is indeed compatible with the background field $\hat{g}_{s\mu\nu}(x)$ that appears in the scalar case (5.75) in \cite{Beneke:2021aip}, in the sense that $\hat{e}\indices{_s^a_\mu}(x)$ and $\hat{g}_{s\mu\nu}(x)$ are related through \eqref{eq:VierbeinDefinition},
\begin{equation}
    \hat{g}_{s\mu\nu}(x) = \hat{e}\indices{_s^a_\mu}(x) \hat{e}\indices{_s^b_\nu}(x)\eta_{ab}\,.
\end{equation}
The second ``Wilson line'' is the soft translation denoted by $R$, which 
defines the fixed-line normal coordinates (FLNC) and is given in (5.60) in \cite{Beneke:2021aip}. It changes the soft GCT transformation of collinear fields from diffeomorphisms at point $x$ to local Poincar\'e transformations at $x_-$.
Combining both Wilson lines, the collinear field redefinition reads 
\begin{align}
    \psi_c&= \left[R W_c^{-1} \tilde{\psi}_c \right],
\label{cfieldreffermion}\\
    e\indices{_c^a_\mu}(x)&=\left[R R\indices{_\mu^\alpha}\left(\left[W\indices{^\rho_\alpha}W\indices{_c^{-1}}\left(\hat{e}\indices{_s^a_\rho}(x)+\tilde{e}\indices{_c^a_\rho}(x)\right) \right]-\hat{e}\indices{_s^a_\alpha}(x)\right)\right]\,,
\label{cfieldrefvierbein}\end{align}
where $\tensor{W}{^\rho_\alpha}$ and $\tensor{R}{_\mu^\alpha}$ are the Jacobians corresponding to $W_c^{-1}$ and $R$, as defined in (5.38) in \cite{Beneke:2021aip}.
The vierbein $\hat{e}\indices{_s^a_\alpha}(x)$ is again the soft background vierbein for the redefined, hatted collinear fields defined in \eqref{eq:VierbeinSplitBG}.
In \eqref{cfieldreffermion}, \eqref{cfieldrefvierbein} the fields on the left-hand side are assumed to be in collinear light-cone gauge. The ``tilde''-fields on the right-hand side transform with the residual ``homogeneous'' GCT symmetry discussed in \cite{Beneke:2021aip}, which takes the form of a local Poincar\'e transformation along the classical trajectory of the energetic particles $x_-^\mu$.
The hatted fields are the emergent soft background fields compatible with this symmetry, and constructed explicitly in Sec.\ \ref{sec:EmergentBG}. 
Let us emphasise the strong formal similarity to gauge theory, despite the very different underlying symmetry.

These ``tilde''-fields, or, more precisely, the collinear gauge-invariant building blocks constructed from these fields, form the starting point of the following discussion treating the local Lorentz transformations.

\section{Wilson lines and field redefinitions for LLT}\label{sec:Wilson}

The next step is to perform the analogous redefinitions also for the local Lorentz transformations.
That is, the hatted fields with homogeneous LLT transformation \eqref{eq:defining_soft_LLT_trafos} must be related to the unhatted full-theory fields via the GCT-redefined tilde-fields.
Since LLT take the form of a standard gauge symmetry one can define the standard Wilson line \eqref{eq:StandardLLTWilsonLine}, and the redefinitions follow the same spirit as in gauge-theory.

The first ingredient is the LLT-analogue of $W_c$, which is used to render collinear fields manifestly gauge-invariant under collinear transformations.
This object can already be identified in the purely-collinear theory, so for the following discussion we set the soft graviton and correspondingly the soft spin-connection and vierbein to zero.

Since the final building block should be invariant under both GCT and LLT, it is convenient to directly consider the
gauge-invariant GCT building block $W_c^{-1}\hat{\psi}_c$.
If one fixes collinear GCT light-cone gauge and focuses only on the local Lorentz symmetry, the candidate Wilson lines become evident and are given by the gauge-theory versions \eqref{eq:GaugeWC}, \eqref{eq:GaugeR} with the replacements $A_\mu \to \Omega_\mu$ and $g_s\to 1$.
For example, the collinear LLT-Wilson line
\begin{equation}\label{eq:SimpleLLTWC}
    W_{c,{\rm LLT}}(x) = \textbf{P}\exp\lp i\int_{-\infty}^0 \de s\: \np \hat{\Omega}_c(x+sn_+)\rp
\end{equation}
transforms as
\begin{equation}\label{eq:SimpleLLTCollTrans}
    W_{c,{\rm LLT}}(x) \to D_{\Lambda_c}(x)W_{c,{\rm LLT}}(x)\,,
\end{equation}
and one can construct a LLT-invariant spinor field simply by taking $W_{c,{\rm LLT}}^{-1}(x)\hat{\psi}_c(x)$.
This Wilson line fixes the local Lorentz analogue of light-cone gauge, namely $\np\hat{\Omega}_c(x) = 0$.
In this gauge, the Wilson line simply becomes unity in the given representation, $W_{c,{\rm LLT}} = \mathds{1}$.

To reinstate the collinear GCT, one considers the GCT-invariant field $[W_c^{-1}\hat{\psi}_c]$.
When performing a local Lorentz transformation of $[W_c^{-1}\hat{\psi}_c]$, one has to keep in mind that the GCT Wilson line $W_c^{-1}$ is a derivative operator that also acts on the Lorentz transformation, but is itself LLT-invariant.
The resulting transformation reads
\begin{equation}
    \lc W_c^{-1}\hat{\psi}_c\rc \to \lc W_c^{-1} D_{\Lambda_c}(x)\hat{\psi}_c\rc = D_{\Lambda_c}(x+\theta_{\rm LC}(x)) \lc W_c^{-1}\hat{\psi}_c\rc\,,
\end{equation}
making use of the product rule (see App. C of \cite{Beneke:2021aip})
\begin{equation}
    \lc W_c^{-1}\phi(x)\psi(x)\rc = \lc W_c^{-1}\phi(x)\rc\lc W_c^{-1}\psi(x)\rc\,,
\end{equation}
and recasting the action of $W_c^{-1}$ as a translation by $\theta_{\rm LC}(x)$. The GCT-invariant field $[W_c^{-1}\hat\psi_c](x)$ does not transform with $D_\Lambda(x)$, but rather with the translated Lorentz-trans\-for\-mation $D_{\Lambda_c}(x+\theta_{\rm LC}(x)).$
Therefore, the analogue of $W_c$ is a Wilson line $V_c$ which moves the transformation from point $x+\theta_{\rm LC}(x)$ to $-\infty$ along the $n_+^\mu$ direction.
Or, in other words, one must translate the Wilson line $W_{c,{\rm LLT}}(x)$.
This translated Wilson line is 
\begin{align}
    \label{eq:collinear_LLT_Wilson_line}
     V_c(x)&\equiv\textbf{P} \exp{\left(i\int_{-\infty}^{0} \de s\: n^\mu_+ \hat{\Omega}_{c\mu} \bigl(x+ sn_+ +\theta_{\rm LC}(x+ sn_+)\bigr)\right)}\nn\\
&=\textbf{P} \exp{\left(i\int_{-\infty}^{0} \de s\: n_+^\mu W\indices{^\rho_\mu}\left[W_c^{-1}\hat{\Omega}_{c\rho} \right](x+sn_+)\right)}\,,
\end{align}
where we pulled out the effect of the translation $x\to x + \theta_{\rm LC}(x)$ in the second line by dressing the field $\hat{\Omega}_{c\mu}$.
Notably, this object $W\indices{^\rho_\mu}[W_c^{-1}\hat{\Omega}_{c\rho}]$ is simply the GCT-invariant spin-connection and reduces to $\hat{\Omega}_{c\mu}$ when fixing GCT light-cone gauge.
One immediately verifies that $V_c(x)$ has indeed the desired transformation
\begin{equation}
    V_c(x) \to D_{\Lambda_c}\bigl(x+\theta_{\rm LC}(x)\bigr)V_c(x)\,,
\end{equation}
and it reduces to the simple \eqref{eq:SimpleLLTCollTrans} in GCT light-cone gauge, where $\theta_{\rm LC}^\mu(x)=0$.
Therefore, the manifestly LLT and GCT gauge-invariant spinor field is defined as
\begin{equation}
    \hat{\Psi}_c \equiv V_c^{-1} \lc W_c^{-1}\hat{\psi}_c\rc\,.
\end{equation}
Note that this Wilson line corresponds to a different gauge condition compared to the naive \eqref{eq:SimpleLLTWC}.
Here, the gauge condition is local Lorentz light-cone gauge of the translated field,\footnote{If one fixes the GCT light-cone gauge first, this translated field agrees with $\hat{\Omega}_c$ and the gauge condition is the standard LLT light-cone gauge.}
\begin{equation}
    \np \hat{\Omega}_c(x+\theta(x)) = \np^\mu W\indices{^\rho_\mu}\lc W_c^{-1} \hat{\Omega}_{c\rho}\rc(x) = 0\,.
\end{equation}

It is straightforward to extend this definition to the case where the soft graviton field is present.
In this case, similar to the GCT-Wilson line $W_c$ discussed in Sec.~5.6.1 in \cite{Beneke:2021aip}, the full effective spin-connection $\hat\Omega(x)=\hat\Omega_c(x)+\hat\Omega_s(x)$, 
including the soft spin-connection 
computed from the soft background vierbein given in \eqref{eq:soft_background_vierbein_symmetric} below, appears in the exponent of $V_c(x)$. 
This is required since all objects are defined with respect to this residual soft background, and the Wilson lines should have the corresponding soft-covariant transformation.\footnote{In QCD, this does not happen since the soft background field $\nm A_s(x_-) \frac{\np^\mu}{2}$ is projected out by the $\np^\mu$ in the integrand.
In gravity, however, there is a non-vanishing component $s_{-+}\np^\mu \nm^\nu/4$ in the background field, and consequently $W_c$ depends also on the soft graviton at the non-linear level.
Unsurprisingly, the same occurs in the LLT context.}

To derive the analogue of $R$ in \eqref{eq:GaugeR}, one can follow the same intuition.
As a first step, it is convenient to fix both collinear LLT and GCT 
light-cone gauges to remove the $W_c$ and $V_c$ Wilson lines.
The candidate Wilson line is then
\begin{equation}\label{eq:SimpleLLTR}
    R_{\rm LLT}(x) = \textbf{P}\exp\lp i\int_0^1 \de s\: (x-x_-)^\mu \Omega_{s\mu}\bigl(x+s(x-x_-)\bigr)\rp\,,
\end{equation}
which transforms as
\begin{equation}
    R_{\rm LLT}(x) \to D_{\Lambda_s}(x)R_{\rm LLT}(x)D_{\Lambda_s}^{-1}(x_-)\,.
\end{equation}
However, note that the full-theory field $\psi_c(x)$ and the GCT-redefined field $\tilde{\psi}_c(x)$ are related by the GCT Wilson line $R$ via
\begin{equation}
    R^{-1}\psi_c(x) = \tilde{\psi}_c(x)\,.
\end{equation}
The full-theory field transforms as $\psi_c(x)\to D_{\Lambda_s}(x)\psi_c(x)$, which implies for the transformation of $\tilde{\psi}_c$
\begin{equation}
    \tilde{\psi}_c(x) \to \lc R^{-1} D_{\Lambda_s}(x)\rc\tilde{\psi}_c(x) = D_{\Lambda_s}(x+\theta_{\rm FLNC}(x))\tilde{\psi}_c(x)\,,
\end{equation}
where $\theta^\mu_{\rm FLNC}(x)$ is the translation parameter to FLNC that defines $R$.
Like the gauge-invariant building block $[W_c^{-1}\hat\psi_c]$ before, also the intermediate field $\tilde{\psi}_c = [R^{-1}\psi_c]$ has a translated LLT.
Therefore, the proper LLT analogue, now denoted as $V_s$, of the GCT Wilson line $R$ must transform as
\begin{equation}
    V_s(x) \to D_{\Lambda_s}\bigl(x+\theta_{\rm FLNC}(x)\bigr) V_s(x) D_{\Lambda_s}^{-1}(x_-)\,,
\end{equation}
and corresponds to a translated version of \eqref{eq:SimpleLLTR}. 
This translated Wilson line can be expressed in terms of $R$ as
\begin{eqnarray}
     V_s&=&\textbf{P} \exp{\left(i\int_{x_-}^{x} \de y^\mu\: \Omega_{s\mu}(y+\theta_{\rm FLNC}(y))\right)}
=\textbf{P} \exp{\left(i\int_{x_-}^{x} \de y^\mu\:  R\indices{^\rho_\mu}\lc R^{-1}\Omega_{s\rho}\rc (y)\right)}\,,\qquad
    \label{eq:soft_LLT_WilsonLine}
\end{eqnarray}
where we again pulled out the translation in the second equality by dressing $\Omega_{s\mu}$ with $R$.
Note that in fixed-line normal coordinates, $R=1$, $\tensor{R}{^\alpha_\mu}=\delta\indices{^\alpha_\mu}$ and the translated Wilson line $V_s$ agrees with \eqref{eq:SimpleLLTR}, as expected.
Here we used that $\theta^\mu_{\rm FLNC}(x_-)=0$, as follows from 
(5.59) in \cite{Beneke:2021aip}.

In summary, the GCT redefinition of the collinear fields via GCT ``Wilson lines'' leads to a local translation of the local Lorentz transformation
$D_{\Lambda}(x) \to D_{\Lambda}(x+\theta_{\rm LC/FLNC}(x))$,
where $\theta_{\rm LC/FLNC}^\mu(x)$ is either the collinear or soft GCT-parameter.
To account for this shift, one employs the translated Wilson lines \eqref{eq:collinear_LLT_Wilson_line}, \eqref{eq:soft_LLT_WilsonLine}.
They can be expressed as standard Wilson lines with the dressed spin-connections in their exponent.
The collinear Wilson line \eqref{eq:collinear_LLT_Wilson_line} is the gauge-theory analogue \eqref{eq:GaugeWC} with the GCT-invariant spin-connection $\tensor{W}{^\rho_\mu}[W_c^{-1}\hat{\Omega}_{\rho}](x)$ instead of $A_{c\mu}(x)$.
For the soft Wilson line \eqref{eq:soft_LLT_WilsonLine}, one finds the $R$ Wilson line \eqref{eq:GaugeR} with the spin-connection in GCT fixed-line gauge $\tensor{R}{^\rho_\mu}[R^{-1}\Omega_{s\rho}](x)$ instead of $A_s(x)$.

The field redefinitions relating the fully homogeneously-transforming hatted fields \eqref{eq:defining_soft_LLT_trafos} to the original fields then read
\begin{equation}\label{eq:redefinition_fields}
\begin{aligned}
    \psi_c&= \left[R V\indices{_s}V\indices{_c^{-1}}W_c^{-1} \left(\hat{\xi}+\hat{\zeta} \right) \right],\\
    e\indices{_c^a_\mu}(x)&=\left[R R\indices{_\mu^\alpha}[V_s]\indices {^a_b}\left(\left[\left[V\indices{_c}\right]\indices{_d^b}W\indices{^\rho_\alpha}W\indices{_c^{-1}}\left(\hat{e}\indices{_s^d_\rho}(x)+\hat{e}\indices{_c^d_\rho}(x)\right) \right]-\hat{e}\indices{_s^b_\alpha}(x)\right)\right]\,,
\end{aligned}
\end{equation}
where the following projection identities hold for $\hat{\psi}_c = 
\hat{\xi}+\hat{\zeta}$:
\begin{equation}
    \hat{\xi}=\frac{\slashed{n}_-\slashed{n}_+}{4}\hat{\xi}\, ,\qquad
    \hat{\zeta}=\frac{\slashed{n}_+\slashed{n}_-}{4}\hat{\zeta}\,.
\end{equation}
In \eqref{eq:redefinition_fields}, $V_s$ and $V_c^{-1}$ denote the LLT Wilson lines in the Dirac spinor representation, $[V]_{\alpha\beta} = [D_{\Lambda_{V}}]_{\alpha\beta}$, whereas $[V_s]\indices{^a_b}$ and $[V_s]\indices{_d^b}$ are in fundamental representation, i.e.\ local Lorentz transformations $[\Lambda_{V}]\indices{^a_b}$, in agreement with the representation of $\psi_c$ and $e\indices{_c^a_\mu}(x)$, respectively.
The fields on the left-hand side of \eqref{eq:redefinition_fields} are evaluated in collinear LLT and GCT light-cone gauge, while the hatted fields on the right-hand side are not gauge-fixed.
Under a soft gauge transformation, the fields on the left-hand side transform with the original, $x$-dependent transformation \eqref{eq:starting_soft_LLT_trafos}, while the hatted fields transform with the $x_-$-dependent ones as in~\eqref{eq:defining_soft_LLT_trafos}. 

\section{Emergent soft background}
\label{sec:EmergentBG}

\subsection{Vierbein in fixed-line normal coordinates}

When inserting the redefinitions \eqref{eq:redefinition_fields} into the Lagrangian, one finds that the soft vierbein $e\indices{_s^a_\mu}(x)$ \eqref{eq:vierbeinsplit} appears only through the combination
\begin{equation}
\label{eq:full:dressed_soft_vierbein}
    \check{e}\indices{_s^a_\mu}(x)=[V_s]\indices{_b^a}R\indices{^\nu_\mu}\left[R^{-1}e\indices{_s^b_\nu}(x)\right]\, .
\end{equation}
The effect of these Wilson lines is that the multipole-expanded soft vierbein
\begin{align}
    e\indices{_s^a_\mu}(x) &= e\indices{_s^a_\mu}(x_-) + (x-x_-)^\alpha\lc \partial_\alpha e\indices{_s^a_\mu}\rc(x_-)\nn\\
    &\quad+ \frac 12 (x-x_-)^\alpha (x-x_-)^\beta \lc \partial_\alpha\partial_\beta e\indices{_s^a_\mu}\rc(x_-) + \mathcal{O}\lp(x-x_-)^3\rp
\end{align}
is re-arranged as
\begin{align}
\label{eq:full_dressed_soft_vierbein_taylor}
    \check{e}\indices{_s^a_\mu}(x)&= A\indices{_s^a_\mu}(x_-)+\left(x-x_-\right)^\alpha B\indices{_s^a_\alpha_\mu}(x_-)\nn\\
    &\quad +\frac 12\left(x-x_-\right)^\alpha\left(x-x_-\right)^\beta C\indices{_s^a_\alpha_\beta_\mu}(x_-)+\mathcal{O}\left(\left(x-x_-\right)^3\right)\,,
\end{align}
where, crucially, the expansion coefficients tranform under LLT with $\Lambda\indices{^a_b}(x_-)$.
To make this residual transformation explicit, one may rewrite \eqref{eq:full_dressed_soft_vierbein_taylor} as
\begin{align}\label{eq:full_dressed_soft_vierbein_taylor_symmetric}
   \check{e}\indices{_s^a_\mu}(x)&=\Xi\indices{^a_b}(x_-)\biggl(\lc\tilde{A}\indices{_s^b_\mu}(x_-)+\left(x-x_-\right)^\alpha \tilde{B}\indices{_s^b_\alpha_\mu}\rc\nn\\
   &\quad+\frac 12\left(x-x_-\right)^\alpha\left(x-x_-\right)^\beta \tilde{C}\indices{_s^b_\alpha_\beta_\mu}+\mathcal{O}\left(\left(x-x_-\right)^3\right)\biggr)\, .
\end{align}
As the LLT properties of \eqref{eq:full:dressed_soft_vierbein} are now  contained in the prefactor $\Xi\indices{^a_b}(x_-)$, this implies that $\tilde{A}\indices{_s^a_\mu}(x_-),$ $\tilde{B}\indices{_s^a_\alpha_\mu}(x_-),$ $\tilde{C}\indices{_s^a_\alpha_\beta_\mu}(x_-)$ and all higher-order structures are completely determined through the GCT redefinition employing fixed-line normal coordinates \cite{Beneke:2021aip}.

These coordinates are constructed such that the coefficient $\tilde{C}\indices{_s^a_\alpha_\beta_\mu}(x_-)$ and all higher-order coefficients in the multipole expansion are expressed in terms of GCT gauge-covariant objects, e.g.~the Riemann tensor $R\indices{^\mu_\nu_\alpha_\beta}(x_-)$ and its covariant derivatives.
However, the first two terms of the multipole expansion, i.e.\ $\tilde{A}\indices{_s^b_\mu}(x_-)$ and $\tilde{B}\indices{_s^b_\alpha_\mu}(x_-)$ inside the square bracket in \eqref{eq:full_dressed_soft_vierbein_taylor_symmetric}, are not completely constrained by the FLNC gauge condition,
and transform non-trivially under GCT.
This residual transformation can be parameterised by two corresponding functions $\varepsilon_\mu(x_-)$ and $\lambda_{\alpha\mu}(x_-)$, see e.g.\ (5.91) in \cite{Beneke:2021aip}.

The soft GCT $R$ Wilson line that moves the field to fixed-line gauge is determined by the parameter $\theta_{\rm FLNC}(x)$, given in (5.59) in \cite{Beneke:2021aip}.
In the definition of $\theta_{\rm FLNC}(x)$, one employs a local inertial frame to relate the metric tensor $g_{s\mu\nu}(x_-)$ to the Minkowski metric via (5.53).
There is no distinguished local inertial frame, and any choice is possible.
The particular choice of a frame has an effect on the precise form of the coefficients appearing in \eqref{eq:full_dressed_soft_vierbein_taylor_symmetric}, but does not change the form of the emergent background metric $\hat{g}_{s\mu\nu}(x)$ (5.76) -- (5.79).

To be consistent with \cite{Beneke:2021aip}, we now adopt the same choice of the local inertial frame, implicitly defined in (5.54).
This means that in the $R$ Wilson line, we employ $\theta_{\rm FLNC}$ as given in (5.59).
We also fix the gauge of the soft vierbein $e\indices{_s^a_\mu}(x)$ on the right-hand side of \eqref{eq:full:dressed_soft_vierbein} to symmetric gauge, which amounts to setting $\Xi\indices{^a_b}(x_-)=\delta\indices{^a_b}$ in \eqref{eq:full_dressed_soft_vierbein_taylor_symmetric}.\footnote{If one wants to consider a different or no gauge-fixing for the soft vierbein, one can keep the explicit factor of $\Xi\indices{^a_b}(x_-)$ and carry this throughout the computations.}
This implies that the local inertial frame chosen for the $R$ Wilson line is aligned with the local inertial system defined by the vierbein $e\indices{_s^a_\mu}(x)$ in symmetric gauge,
and one obtains the particularly simple result for the left-hand side of \eqref{eq:full:dressed_soft_vierbein} given below.\footnote{Such a simple result is obtained whenever the local inertial frame of the FLNC coordinates and the one defined by the vierbein are aligned.}

\subsection{Background vierbein in symmetric gauge}

Using these Wilson lines and the combination $\check{e}\indices{_s^a_\mu}(x)$ \eqref{eq:full:dressed_soft_vierbein}, one can identify the residual soft background vierbein of SCET gravity, i.e.~the analogue of the emergent background metric $\hat{g}_{s\mu\nu}(x)$, given in (5.76) -- (5.79) in \cite{Beneke:2021aip}.
This residual background corresponds to the components of the metric tensor, or vierbein, that are \emph{not constrained} by the gauge-condition implemented via the Wilson lines.
To identify this background, the object $\check{e}\indices{_s^a_\mu}(x)$ is  split into a manifestly gauge-covariant term $\mathfrak{e}\indices{_s^a_\mu}(x)$, which is expressed via the Riemann tensor, and the residual background field $\hat{e}\indices{_s^a_\mu}(x)$, in complete analogy to (5.75) in \cite{Beneke:2021aip} for the metric tensor.
This split reads
\begin{equation}\label{eq:VierbeinSplitBG}
   \check{e}\indices{_s^a_\mu}(x)\equiv    \hat{e}\indices{_s^a_\mu}(x)+ \mathfrak{e}\indices{_s^a_\mu}(x)\,,
\end{equation}
and the background $\hat{e}\indices{_s^a_\mu}(x)$ corresponds to the first two terms of \eqref{eq:full_dressed_soft_vierbein_taylor_symmetric}, while the gauge-covariant term $\mathfrak{e}\indices{_s^a_\mu}(x)$ contains the third term in \eqref{eq:full_dressed_soft_vierbein_taylor_symmetric}, as well as \emph{all} higher-order terms due to the multipole expansion.

Fixing symmetric gauge for the vierbein $e\indices{_s^a_\mu}(x)$ and taking the $R$ Wilson line from \cite{Beneke:2021aip} amounts to setting $\Xi\indices{^a_b}(x_-)=\delta\indices{^a_b}$ in \eqref{eq:full_dressed_soft_vierbein_taylor_symmetric}.
This completely fixes the remaining local Lorentz symmetry, and the redefined vierbein \eqref{eq:full:dressed_soft_vierbein} may be expressed solely via the symmetric vierbein  $[e_{\rm sym}]\indices{^a_\mu}(x)$, its inverse $[E_{\rm sym}]\indices{^\mu_a}(x)$, and their derivatives. 
From now on, we drop the subscript indicating the symmetric vierbein, i.e.\ $[e_{\rm sym}]\indices{^a_\mu}(x_-)=e\indices{^a_\mu}(x_-)$.

To evaluate the expression for the redefined vierbein \eqref{eq:full:dressed_soft_vierbein}, one has to compute the soft GCT Wilson line $R$ given in (5.60) and the parameter $\theta_{\rm FLNC}^\mu(x)$ (5.59) in \cite{Beneke:2021aip} as well as the soft Wilson line $V_s$ \eqref{eq:soft_LLT_WilsonLine}.
Furthermore, the result may  be obtained in a closed form by employing the tetrad postulate $\left(\partial_\mu -i\Omega\indices{_s_\mu}(x)\right)e\indices{_s^a_{\nu}}(x)=e\indices{_s^a_\rho}(x)\Gamma^\rho_{s\mu\nu}(x)$ for the soft vierbein $e\indices{_s^a_\mu}(x)$, which takes the place of the metric-compatibility constraints (5.33) in \cite{Beneke:2021aip}.
For the first two terms of \eqref{eq:full_dressed_soft_vierbein_taylor_symmetric}, i.e.\ the background vierbein $\hat{e}\indices{_s^a_\mu}(x)$, this leads to
\begin{equation}
    \hat{e}\indices{_s^a_\mu}(x)=
    \lp\delta^a_{\perp\mu}+ 
    \frac{n_{-\mu}}{2}\np^a + \frac{n_{+\mu}}{2}\lp  e\indices{_s^a_-}(x_-)+y^\beta \omega\indices{_-^a_\beta}(x_-)\rp\rp\,,\label{eq:soft_background_vierbein_symmetric}
\end{equation}
where we defined $y^\beta \equiv (x-x_-)^\beta$, and $\omega\indices{_-^b_\beta}(x_-)$ is constructed from $e\indices{_s^a_\mu}(x)$ through
\begin{equation}\label{eq:SoftSpinConnection}
\omega\indices{_\mu_a_b}(x) = g_{s\nu\rho}(x)\tensor{E}{_s^\rho_{a}}(x)\lp \partial_\mu \tensor{E}{_s^\nu_b}(x) + \Gamma^\nu_{s\mu\lambda}(x)E\indices{_s^\lambda_b}(x)\rp\,.
\end{equation}
Here, $\Gamma^\rho_{s\nu\mu}(x)$ is the Christoffel symbol constructed from the soft metric tensor $g_{s\mu\nu}(x)$.
In the following, we drop the arguments of soft fields if they are evaluated at $x_-$, i.e.~$e\indices{_s^a_\mu}(x_-)\equiv e\indices{_s^a_\mu}$.

While both fields $e\indices{_s^a_-}$, $\omega\indices{_-^a_\beta}$, originate from the same ``full-theory'' soft background $e\indices{_s^a_\mu}(x)$, they are evaluated at $x_-^\mu$ and should be interpreted as \emph{independent} gauge fields when viewed from the EFT perspective, as $\omega_{\mu a\beta}(x_-)$ contains information about the transverse derivative of $e\indices{_s^a_\mu}(x)$ that are not enclosed in $e\indices{_s^a_\mu}(x_-)$.

For explicit computations, it is advantageous to perform the weak-field expansion also for the soft background $g_{s\mu\nu}(x)$.
Using the expansion of the symmetric vierbein and spin-connection,
\begin{align}
    \label{eq:SoftVierbein}
     e\indices{_s^a_-} &= \nm^a+\frac{1}{2}s\indices{^a_-}-\frac{1}{8} s\indices{_-_\beta}s\indices{^\beta^a} + \mathcal{O}(\lambda^6)\,,\\
    \omega_{-ab}
    &= -\frac 12 (\partial_{a}s_{b-} - \partial_b s_{a-}) + \mathcal{O}(\lambda^6)\,,\label{eq:SoftSpinConnectionSym}
\end{align}
respectively,
one finds for the residual background vierbein \eqref{eq:soft_background_vierbein_symmetric} to $\mathcal{O}(\lambda^4)$
\begin{equation}\label{eq:BGVierbeinWeakField}
    \hat{e}\indices{_s^a_\mu}(x) =\lp\delta \indices{^a_\mu}+\frac{1}{4}n \indices{_+_\mu}s\indices{^a_-}-\frac{1}{4} n \indices{_+_\mu} y^\rho (\partial^a s_{\rho-} - \partial_\rho s^a_-) -\frac{1}{16} n \indices{_+_\mu}s\indices{_-_\beta}s\indices{^\beta^a}\rp+\mathcal{O}(\lambda^5)\,.
\end{equation}
The gauge-covariant part $\mathfrak{e}\indices{_s^a_\mu}(x)$ reads
\begin{equation}\label{eq:GaugeCovariantVierbein}
   \mathfrak{e}\indices{_s^a_\mu}(x)= -\frac{1}{6}\left( x-x_-\right) ^\rho \left( x-x_-\right) ^\alpha \lp 
   R \indices{^a_\rho_\mu_\alpha}+n_{+\mu} R \indices{^a_\rho_-_\alpha}\rp+\mathcal{O}(\lambda^5)\,.
\end{equation}

Note that while the expression $\check{e}\indices{_s^a_\mu}(x)$ is constructed from the symmetric vierbein, neither the background $\hat{e}\indices{_s^a_\mu}(x)$ nor the gauge-covariant terms $\mathfrak{e}\indices{_s^a_\mu}(x)$ are symmetric themselves, e.g.\ only the $\np^\mu$-component of $\hat{e}\indices{_s^a_\mu}(x)$ is non-trivial, as one would expect since the LLT Wilson line employed in \eqref{eq:full:dressed_soft_vierbein} corresponds to fixed-line gauge and not to symmetric gauge.

The background vierbein $\hat{e}\indices{_s^a_\mu}(x)$ also generates a metric tensor $\hat{g}_{s\mu\nu}(x)$ via
\begin{equation}
    \hat{g}_{s\mu\nu}(x) = \hat{e}\indices{_s^a_\mu}(x)\hat{e}\indices{_s^b_\nu}(x)\eta_{ab}\,.
\end{equation}
Upon inserting \eqref{eq:soft_background_vierbein_symmetric}, one finds
\begin{align}
    \hat{g}_{s\mu\nu}(x) &= \eta_{\perp\mu\nu} + 
    \frac{n_{+\nu}}{2}
    \lp
     e\indices{_s^\alpha_-}+y^\beta \omega\indices{_-^\alpha_\beta}\rp
    \left(\eta\indices{_\perp_\alpha_\mu}+\frac{n\indices{_-_\mu}n\indices{_+_\alpha}}{2}\right)\nn\\
    \label{eq:VierbeinBGMetric}
    &\quad
    +\frac{n_{+\mu}}{2}
    \lp
     e\indices{_s^\alpha_-}+y^\beta \omega\indices{_-^\alpha_\beta}\rp
    \left(\eta\indices{_\perp_\alpha_\nu}+\frac{n\indices{_-_\nu}n\indices{_+_\alpha}}{2}\right)\\
    &\quad+
    \frac{n_{+\mu}n_{+\nu}}{4}\eta_{\alpha\beta}
    \lp
     e\indices{_s^\alpha_-}+y^\rho \omega\indices{_-^\alpha_\rho}\rp
    \lp e\indices{_s^\beta_-}+y^\sigma \omega\indices{_-^\beta_\sigma}\rp\,.\nn
\end{align}

\subsection{Emergent background metric}

One should now compare the result \eqref{eq:VierbeinBGMetric} to the background metric $\hat{g}_{s\mu\nu}(x)$ that emerged in the scalar scenario \cite{Beneke:2021aip}.
First, note that the scalar-case metric can be written as
\begin{align}
    \hat{g}_{s\mu\nu}(x) &= \eta_{\perp\mu\nu} + \frac{n_{+\mu}n_{+\nu}}{4}\hat{g}_{s--}(x) + \frac{n_{+\nu}}{2}\hat{g}_{s\mu_\perp-}(x) + \frac{n_{+\mu}}{2}\hat{g}_{s\nu_\perp-}(x)\nn\\
    &\quad + \lp \frac{n_{+\mu}n_{-\nu}}{4} + \frac{n_{+\nu}n_{-\mu}}{4}\rp \hat{g}_{s+-}(x)\,,\label{eq:ScalarMetricFull}
\end{align}
where the individual components are given in (5.76) -- (5.79) in \cite{Beneke:2021aip}.
One can similarly decompose this into vierbeins using the standard relations \eqref{eq:VierbeinDefinition}
\begin{equation}\label{eq:ScalarMetricDecompDefinition}
    \hat{g}_{s\mu\nu}(x) \equiv \hat{e}\indices{_{\hat{g}}^a_\mu}(x)\hat{e}\indices{_{\hat{g}}^b_\nu}(x)\eta_{ab}\,.
\end{equation}
Due to the specific form in \eqref{eq:ScalarMetricFull}, one sees that only $\delta\indices{_\perp^a_\mu}$ and the component $\hat{e}\indices{_{\hat{g}}^a_-}(x)$ are necessary to describe the metric tensor as
\begin{align}
    \hat{g}_{s\mu\nu}(x) =\Bigl\{ &\delta\indices{_\perp^a_\mu}\delta\indices{_\perp^b_\nu} + \frac{n_{+\mu}n_{+\nu}}{4}\hat{e}\indices{_{\hat{g}}^a_-}(x)\hat{e}\indices{_{\hat{g}}^b_-}(x)
    + \frac{n_{+\nu}}{2}\hat{e}\indices{_{\hat{g}}^b_-}(x)\left(\delta\indices{_\perp^a_\mu}+\frac{n\indices{_-_\mu}n\indices{_+^a}}{2}\right)  \nn \\  & +\frac{n_{+\mu}}{2}\hat{e}\indices{_{\hat{g}}^a_-}(x)\left(\delta\indices{_\perp^b_\nu}+\frac{n\indices{_-_\nu}n\indices{_+_b}}{2}\right)\Bigr\}\, \eta\indices{_a_b}\,,\label{eq:ScalarMetricVierbein}
\end{align}
By comparing the individual terms in \eqref{eq:ScalarMetricVierbein} with \eqref{eq:ScalarMetricFull}, using the explicit form of $\hat{g}_{s\mu\nu}(x)$ given in (5.76) -- (5.79) in \cite{Beneke:2021aip}, one can identify
\begin{equation}\label{eq:ScalarVierbein}
    \hat{e}\indices{_{\hat{g}}^\alpha_-}(x) = e\indices{_{[4]}_-^\alpha} + (x-x_-)^\beta[\Omega_{[4]-}]\indices{_{\beta}^\alpha}\,,
\end{equation}
where $e\indices{_{[4]}_-^a}$ and $[\Omega_{[4]-}]\indices{_{\beta}^a}$ are defined in (5.53) and (5.69) in \cite{Beneke:2021aip}, respectively.
Comparing this expression to \eqref{eq:VierbeinBGMetric}, one sees that this background metric precisely agrees with the one generated by the vierbein $\hat{e}\indices{_s^a_\mu}(x)$ computed in the previous section, as it should be.
This is also evident in the weak-field expansion, where, using
(5.80), (5.81) in \cite{Beneke:2021aip}, one finds 
\begin{align}
    e\indices{_{[4]}_-^\alpha} &= \delta\indices{_-^\alpha} + \frac 12 s\indices{_-^\alpha} - \frac{1}{8}s_{-\beta}s^{\beta\alpha} + \mathcal{O}(\lambda^6)\,,\\
    [\Omega_{[4]-}]\indices{_{\alpha\beta}} &= -\frac 12(\partial_\alpha s_{\beta-} - \partial_\beta s_{\alpha-}) + \mathcal{O}(\lambda^6)\,,
\end{align}
in complete agreement with the result \eqref{eq:BGVierbeinWeakField} for $\hat{e}\indices{_s^a_\mu}(x)$ after identifying $e\indices{_{[4]}_-^\alpha}\to e\indices{_s^a_-}$ and $[\Omega_{[4]-}]_{\alpha\beta} \to \omega_{-\alpha\beta}$.
In other words, when starting from a fermionic theory and using the redefinitions for collinear fields determined in Sec.~\ref{sec:Wilson}, one finds \emph{the same emergent background metric} that is also obtained in the scalar theory, where no LLT are present. 
This is to be expected, since the local Lorentz symmetry is not a physical symmetry and therefore must not affect the residual soft graviton gauge symmetry.
Regardless, this is an important consistency check and shows that the residual metric field is not affected by the additional LLT redefinitions.

This puts the integer-spin and half-integer spin theories on the same footing:
both theories can be described in terms of the same emergent background field $\hat{g}_{s\mu\nu}(x)$ or $\hat{e}\indices{_s^a_\mu}(x)$, which are linked by the standard vierbein definition \eqref{eq:VierbeinDefinition}.

\subsection{Covariant derivative}

In the scalar theory \cite{Beneke:2021aip}, the soft background gives rise to a soft-covariant derivative, which contains two gauge fields, where one of them couples to momentum and the other to the orbital angular momentum.
This derivative arises from the observation that in the kinetic term, one can write
    \begin{equation}
        \hat{g}_s^{\mu\nu}(x)\partial_\mu\varphi\partial_\nu\varphi
        = \partial_{\mu_\perp}\varphi\partial^{\mu_\perp}\varphi + \np\partial\varphi\nm D_{s, \varphi}\varphi\,,
    \end{equation}
since the background metric $\hat{g}_{s\mu\nu}(x)$ is non-trivial only for $\hat{g}_{s\mu-}(x)$.
This defines the covariant derivative through the inverse of \eqref{eq:ScalarMetricFull} as\footnote{Notice that there is a typo in (5.84) in \cite{Beneke:2021aip} in the prefactor of $\omega_{-\alpha\beta}$: the correct expression is \eqref{eq:ScalarMetricCovDev} here.}
\begin{align}\label{eq:ScalarMetricCovDev}
    \nm D_{s,\varphi} &= \frac{1}{4}\hat{g}_s^{--}(x)\np\partial + \hat{g}_{s}^{\mu_\perp-}(x)\partial_{\mu_\perp} +\frac 12 \hat{g}_s^{+-}(x)\nm\partial\nn\\
    &=\partial_--\frac{1}{2}s\indices{_-^\mu}\partial_\mu+\frac{1}{8} s\indices{_-_\alpha}s\indices{^\alpha^\mu}\partial_\mu+\frac{1}{8}s\indices{_-_+}s\indices{^\mu_-}\partial_\mu-\omega\indices{_-^\mu_\rho}\left(x-x_-\right)^\rho\partial_\mu+\mathcal{O}(\lambda^5)\,.
\end{align}
Using the decomposition into the vierbein \eqref{eq:ScalarMetricVierbein}, one sees that this derivative can equivalently be written as
\begin{equation}\label{eq:ScalarVierbeinCovDev}
    \nm D_{s,\varphi} = \hat{E}\indices{_s^\mu_-}(x)\partial_\mu\,,
\end{equation}
 in terms of the inverse vierbein $\hat{E}\indices{_s^\mu_-}(x)$.
This inverse is obtained by inverting \eqref{eq:soft_background_vierbein_symmetric}, or equivalently \eqref{eq:ScalarVierbein}, via \eqref{eq:VierbeinDefinition}, resulting in
\begin{equation}
\label{eq:bckground_vierbein}
\hat{E}\indices{_s^\mu_-}(x) = \nm^\mu -\frac{1}{2}s\indices{_-^\mu} + \frac{1}{8} s\indices{_-_\alpha}s\indices{^\alpha^\mu} +\frac{1}{8}s\indices{_-_+}s\indices{^\mu_-} - \omega\indices{_-^\mu_\rho}\left(x-x_-\right)^\rho +\mathcal{O}(\lambda^5)\,.
\end{equation}
Inserting this into the covariant derivative \eqref{eq:ScalarVierbeinCovDev} immediately yields \eqref{eq:ScalarMetricCovDev}.
In this covariant derivative \eqref{eq:ScalarMetricCovDev}, one observes that the spin-connection $\omega_{-\alpha\beta}$ couples to the orbital angular momentum operator\footnote{In \cite{Beneke:2021aip}, $L^{\mu\nu}$ was defined without the factor $i$ to be consistent with conventions employed in the soft theorems.}
\begin{align}
   L^{\mu\nu} = i(x-x_-)^\mu\partial^\nu -i (x-x_-)^\nu\partial^\mu
\end{align}
 of the scalar field, as seen in the last term in \eqref{eq:ScalarMetricCovDev}, using the antisymmetry $\omega\indices{_-_\rho_\mu}=-\omega\indices{_-_\mu_\rho}$ to rewrite
    \begin{equation}
       -\omega\indices{_-^\mu_\rho} \left(x-x_-\right)^\rho\partial_\mu=-\frac{i}{2}\omega\indices{_-_\mu_\rho}L\indices{^\mu^\rho}\,.
    \end{equation}
Consistency of the framework implies that in the fermionic theory, the orbital angular momentum should be promoted to the full angular momentum including spin. 

Indeed, for fermions the soft-covariant derivative due to the soft background combines with the already-present local Lorentz-covariant derivative, and leads to the object 
    \begin{align}\label{eq:covder}
       i\nm D_s
      =\hat{E}\indices{_s^\mu_-}(x)i\hat{D}\indices{_s_\mu}
      =in_-D\indices{_{s,\varphi}} + \hat{\Omega}\indices{_s_-}(x_-)\, +\mathcal{O}(\lambda^5)\,,
    \end{align}
where the last term, $\hat{\Omega}\indices{_s_-}(x_-)$, is the spin-connection
in spin-$\frac{1}{2}$ representation constructed from $\hat{e}\indices{_s^a_\mu}(x)$ given in \eqref{eq:soft_background_vierbein_symmetric} through its definition \eqref{eq:defining_spin_connection}. 
It stems from the local Lorentz-covariant derivative 
\begin{equation}
i\hat{D}\indices{_s_\mu} \equiv i\partial_\mu + \hat{\Omega}\indices{_s_\mu}(x)
\end{equation}
present in the Lagrangian of fields with spin.

This term is new compared to the scalar result (5.84) in \cite{Beneke:2021aip}, and demands closer inspection.
The spin-connection $\hat{\Omega}_{s\mu}(x)$ contains the spin generator $\Sigma^{ab}$ as
\begin{align}
        \hat{\Omega}_{s\mu}(x) = \frac 12\hat{\omega}\indices{_\mu_a_b}(x)\Sigma\indices{^a^b}&=\frac 12\Sigma\indices{^a^b}\left[\partial_-\hat{e}\indices{_s^c_a}\eta\indices{_c_b}-\frac{1}{2}\partial_a\hat{g}\indices{_s_-_b}+\frac{1}{2}\partial_b\hat{g}\indices{_s_-_a} \right]+\mathcal{O}\left(\lambda^6\right) \nn\\
        &=\frac 12 \Sigma\indices{^a^b}\left(\delta\indices{_b^\rho}-\frac{1}{2}n\indices{_+_b}n\indices{_-^\rho} \right)\omega\indices{_-_a_\rho}\left(\delta\indices{_a^\mu}+\frac{1}{2} n\indices{_+_a}n\indices{_-^\mu}\right)+\mathcal{O}\left(\lambda^6\right)\nonumber\\
        &=\frac 12\omega\indices{_-_\mu_\rho} \Sigma^{\mu\rho}\,,
    \end{align}
where the last equality only holds when wedged between $\slashed{n}_+$ and $\slashed{n}_-$, as will be the case in the Lagrangian \eqref{eq:L_0}.
Hence, the covariant derivative for the spin-$\frac{1}{2}$ field reads
  \begin{equation}
      \nm D_s=\partial_- -\frac{1}{2}s\indices{_-^\mu}\partial_\mu+\frac{1}{8} s\indices{_-_\alpha}s\indices{^\alpha^\mu}\partial_\mu
      + \frac{1}{8}s_{-+}s\indices{_-^\mu}\partial_\mu
      -\frac{i}{2}\omega\indices{_-_\mu_\rho}\left(L\indices{^\mu^\rho}+\Sigma\indices{^\mu^\rho} \right) + \mathcal{O}(\lambda^5)\,,
    \end{equation}
    with the \emph{full} angular momentum operator $L\indices{^\mu^\nu}+\Sigma\indices{^\mu^\nu}$, as was expected.
    This illuminates how the subleading gauge field $\omega_{-\mu\nu}$ couples to the angular momentum:
    the coupling to the orbital piece is due to the factor $(x-x_-)^\mu$ originating from the light-front multipole expansion, and is present for any field \emph{regardless of spin}, since the multipole expansion is a completely generic feature of the SCET construction.
    
    The spin-dependent term, in this case the coupling to $\Sigma^{\mu\nu}$, originates from the original spin-connection, i.e.~from the spin-dependent terms of the full theory.
    All terms combine into one soft-covariant derivative due to the special form of the residual background field $\hat{g}_{s\mu\nu}(x)$, which can be expressed purely in terms of the vierbein $\hat{e}\indices{_s^a_\mu}(x)$.
    The construction is now completely general and can be performed for any spin-$s$ representation, where the full angular momentum should appear in the same fashion.
    The appearance of the subleading gauge field $\omega_{-\mu\nu}$ coupling to the orbital angular momentum in the scalar case is therefore not an accident, but consistently extends to incorporate the full angular momentum for fields with spin.


\section{Building blocks}

In the Lagrangian and the $N$-jet operators, it is convenient to employ collinear gauge-invariant building blocks.
These building blocks are given by 
\begin{equation}
 \label{eq:invariant_building:blocks_full_h}
    \hat{\mathfrak{X}}_c=V_c^{-1}\left[W_c^{-1}\hat{\xi}\right]\,,\quad \hat{\mathfrak{h}}\indices{_\mu_\nu}=W\indices{_c^\alpha_\mu}W\indices{_c^\beta_\nu}\Bigl[W_c^{-1}\hat{h}\indices{_\alpha_\beta}\Bigr]+ W\indices{_c^\alpha_\mu}W\indices{_c^\beta_\nu}\Bigl[W_c^{-1}\hat{g}\indices{_s_\alpha_\beta}\Bigr]-\hat{g}\indices{_s_\mu_\nu}\,,
\end{equation}
where we suppressed the argument $x$ for all appearing fields.
These composite operators correspond to fields in collinear light-cone gauge for both diffeomorphisms and local Lorentz transformations. 
Note that $\hat{\mathfrak{h}}\indices{_\mu_\nu}$ has two Greek indices, so it is already manifestly invariant under LLT.
In addition, one can use the graviton equation of motion \cite{Beneke:2022pue} to eliminate the subleading components $\hat{\mathfrak{h}}\indices{_-_-}$ and $\hat{\mathfrak{h}}\indices{_\bot_-}$ in terms of the physical transverse components $\hat{\mathfrak{h}}_{\perp\perp}$. 
These building blocks are covariant under the emergent soft gauge transformation.

Likewise, any purely-soft operator is manifestly collinear gauge-invariant by definition.
Therefore, soft building blocks need no redefinition and one can simply take gauge-covariant combinations, like the covariant derivative $\nm D_s$ or the Riemann tensor $\tensor{R}{^\mu_{\nu\alpha\beta}}$.
Similar to the scalar case \cite{Beneke:2021aip}, as well as QCD \cite{Beneke:2017ztn}, one can use the matter and graviton equations of motion to eliminate the soft-covariant derivative $\nm D_s$ as a possible soft building block in favour of collinear and other soft gauge-covariant building blocks.

\section{Soft-collinear gravity Lagrangian for fermions}

We now proceed to construct the Lagrangian to $\mathcal{O}(\lambda^2)$ for the Dirac spinor.
The starting point is the curved-space action \eqref{eq:DiracAction}, where one first inserts the decompositions \eqref{eq:metric_split}, \eqref{eq:vierbeinsplit} and \eqref{eq:SpinConnectionSplit} of the gravitational field, decomposing the gravitational field into its soft and collinear modes.
The collinear fields have to be redefined according to \eqref{eq:redefinition_fields}.
Then it is convenient to work in terms of the collinear gauge-invariant building blocks $\hat{\mathfrak{X}}_c$, $\hat{\mathfrak{h}}_{\mu\nu}$ defined in \eqref{eq:invariant_building:blocks_full_h}.
Furthermore, for the intermediate steps, one can introduce the collinear LLT and GCT gauge-invariant $\hat{\mathfrak{Z}}_c$ defined as
\begin{equation}
\hat{\mathfrak{Z}}_c = V_c^{-1}\lc W_c^{-1}\hat{\zeta}\rc\,,
\end{equation}
the analogue of $\hat{\mathfrak{X}}_c$ for $\hat\zeta$.
These components, $\hat\zeta$ or $\hat{\mathfrak{Z}}_c$, are the subleading components in $\lambda$ of the spinor field $\hat\psi_c$ from \eqref{eq:redefinition_fields}, so one should integrate out these fields.
At this point, one can also add the soft fermion field $q(x)$ by modifying \eqref{eq:redefinition_fields} as
\begin{equation}\label{eq:softquark}
    \hat{\psi}_c(x) \to \hat{\xi}(x)+\hat{\zeta}(x) + \Bigl[ W_c V_c V_s^{-1}\lc R^{-1}q(x)\rc\Bigr]\,.
\end{equation}
The additional Wilson lines in front of the soft quark are necessary to implement the correct behaviour under collinear gauge transformations \eqref{eq:GCTcoll}, \eqref{eq:defining_collinear_LLT_trafos}.

It is now straightforward to derive the Lagrangian. The computation proceeds in the same way as the one presented in Sec.~5.6.2 in \cite{Beneke:2021aip}.
The key insight is that the soft Wilson lines $R$, $V_s$ appear in precise combinations to ``dress'' all appearing soft fields in the Lagrangian.
These dressed fields correspond to soft fields in fixed-line gauge, which have a decomposition into the background field and manifestly gauge-covariant parts.
For the vierbein, this split is given in \eqref{eq:VierbeinSplitBG}.
Therefore, one can express the Lagrangian in terms of the emergent soft background field, which gets absorbed into the covariant derivative, and further manifestly gauge-covariant terms that feature the Riemann tensor, just as in the scalar case (5.116) in \cite{Beneke:2021aip}.

In the following we explain how these objects arise for \eqref{eq:DiracAction} with $E\indices{^\mu_a}(x) \to E\indices{_s^\mu_a}(x)$, $\Omega_\mu(x)\to \Omega_{s\mu}(x)$ and $q(x)=0$, i.e.~setting the collinear graviton and soft quark field to zero for illustration, and provide the full Lagrangian in the end.
The starting point is
\begin{equation}
    \mathcal{L} = \sqrt{-g_s(x)}\;\overline{\psi}_cE\indices{_s^\mu_a}(x)\gamma^a (i\partial_\mu +\Omega_{s\mu}(x))\psi_c\,,
\end{equation}
which we take to be in collinear light-cone gauge.
Here, we indicate the explicit $x$-dependence of the soft fields, as they are not yet multipole expanded.
One now inserts the redefinitions \eqref{eq:redefinition_fields} for the spinor field and obtains
\begin{equation}
    \mathcal{L} = \sqrt{-g_s(x)}\lc R V_s \hat{\Psi}_c\rc^\dagger \gamma^0 E\indices{_s^\mu_a}(x)\gamma^a (i\partial_\mu +\Omega_{s\mu}(x))\lc R V_s \hat{\Psi}_c\rc\,,
\end{equation}
where $\hat{\Psi}_c = \hat{\mathfrak{X}}_c + \hat{\mathfrak{Z}}_c$.
To simplify this expression, one first pulls out the $R$-Wilson line using the identities of \cite{Beneke:2021aip}, App.~C.
This yields (up to an irrelevant total derivative)
\begin{equation}\label{eq:RewritingTricks}
    \mathcal{L} = \det(\underline{R})\lc R^{-1}\sqrt{-g_s}\rc\!  R\indices{_\mu^\alpha} \lc R^{-1}E\indices{_s^\mu_a}\rc\!(x) \lp V_s\hat{\Psi}_c\rp^\dagger \gamma^0\gamma^a \bigl(i\partial_\alpha + R\indices{^\nu_\alpha}\lc R^{-1}\Omega_{s\nu}\rc\bigr)V_s\hat{\Psi}_c\,.
\end{equation}
One can already identify the GCT-dressed metric determinant in the first square bracket, the dressed inverse vierbein in the second square bracket and the dressed spin-connection in the third square bracket. All three objects come together with the respective Jacobians $\det(\underline{R})$, $\tensor{R}{_\mu^\alpha}$ and $\tensor{R}{^\nu_\alpha}$ according to their representation.
The metric tensor $\det(\underline{R})\lc R^{-1}\sqrt{-g_s}\rc$ can already be expanded in the background metric $\hat{g}_{s\mu\nu}(x)$ and the Riemann tensor terms using (5.115) in \cite{Beneke:2021aip}. 
Only the leading term is relevant for the Lagrangian to $\mathcal{O}(\lambda^2)$, so one can replace it by $\sqrt{-\hat{g}_s}(x)$ for our purposes.

Next, consider the $V_s$ Wilson line. Here, one can rewrite
\begin{align}
    \lp V_s\hat{\Psi}_c\rp^\dagger &\gamma^0 \gamma^a \bigl(i\partial_\alpha + R\indices{^\nu_\alpha}\lc R^{-1}\Omega_{s\nu}\rc\!(x)\bigr) V_s\hat{\Psi}_c\nn\\
    &= \overline{\hat{\Psi}}_c V_s^{-1}\gamma^a V_s \lp i \partial_\alpha + V_s^{-1} R\indices{^\nu_\alpha}\left[R^{-1}\Omega_{s\nu}\right]\!(x)V_s + iV_s^{-1}\lc \partial_\alpha, V_s\rc\rp \hat{\Psi}_c\,.
\end{align}
The second and third term inside the bracket form the LLT-dressed spin-connection.
The Wilson lines in spinor-representation can be rearranged into $[V_s]\indices{^a_b}$ in vector representation using the $\gamma$-matrix property
\begin{equation}
    V^{-1}_{s}\gamma^aV_{s}=[V_s]\indices{^a_b}\gamma^b\,.
\end{equation}
The resulting $[V_s]\indices{^a_b}$ then multiplies the vierbein, and one finds
\begin{align}
\mathcal{L} &= \det(\underline{R})\lc R^{-1}\sqrt{-g_s}\rc\!(x) [V_s]\indices{^a_b} R\indices{_\mu^\alpha}\lc R^{-1} E\indices{_s^\mu_a}\rc\!(x)\nn\\
&\quad\times\overline{\hat{\Psi}}_c \gamma^b (i\partial_\alpha + V_s^{-1}iR\indices{^\nu_\alpha}\lc R^{-1}\Omega_{s\nu}\rc\!(x) V_s + iV_s^{-1}[\partial_\alpha,V_s])\hat{\Psi}_c\,.
\end{align}
For these composite products of Wilson lines and fields, one can now employ \eqref{eq:full:dressed_soft_vierbein}, \eqref{eq:VierbeinSplitBG} to express the result in terms of the emergent background vierbein $\hat{e}\indices{_s^a_\mu}(x)$ and the Riemann tensor terms.

Performing these manipulations of the Wilson lines for all terms, including the collinear graviton and the soft quark, one finds the compact result
\begin{align}
 \label{eq:Lagrangian_with_redefined_fieldsINV}
\mathcal{L} &= \sqrt{-\det\left(\hat{g}\indices{_s_\mu_\nu}(x)+\hat{\mathfrak{h}}\indices{_\mu_\nu}(x)\right)} \nonumber \\ 
    &\quad\times \left(\overline{\Hat{\mathfrak{X}}}(x)+\overline{\Hat{\mathfrak{Z}}}(x)+\lc R^{-1}\overline{q}(x)\rc V_s   \right)
    \gamma^a\left(\hat{E}\indices{_s^\rho_a}(x) + \hat{\mathsf{E}}\indices{_c^\rho_a}+\mathfrak{E}\indices{_s^\rho_a}(x)\right)\nonumber \\
    &\quad\times i\hat{D}_\rho(x)\left( \hat{\mathfrak{X}}(x)+\hat{\mathfrak{Z}}(x)+V_s^{-1}\lc R^{-1} q(x)\rc\right) +\mathcal{O}(\lambda^7)\,,
\end{align}
where we dropped the Riemann-terms from the metric determinant, as they do not contribute to the given order in $\lambda$.
Here, a number of new objects appear.
The first one is the covariant derivative $\hat{D}_\mu$, which is defined as
\begin{equation}\label{eq:ConstructionInvCovDev}
    \hat{D}_\mu = \hat{D}_{s\mu} - i\hat{\Omega}_\mu(x)\,.
\end{equation}
where $\hat{\Omega}_\mu(x)$ is the spin-connection \eqref{eq:defining_spin_connection} constructed from $\hat{E}\indices{_s^\rho_a}(x) + \hat{\mathsf{E}}\indices{_c^\rho_a}(x)$ and the corresponding metric tensor $\hat{g}_{s\mu\nu}(x)+\mathfrak{h}_{\mu\nu}(x)$.
Here, $\hat{E}\indices{_s^\rho_a}(x)$  is
 the inverse soft background vierbein constructed from \eqref{eq:soft_background_vierbein_symmetric}, with weak-field expansion  given in \eqref{eq:bckground_vierbein},
and the collinear gauge-invariant vierbein $\hat{\mathsf{E}}\indices{_c^\mu_a}(x)$ is determined by applying the collinear GCT Wilson lines $W_c$ to the vierbein $\hat{E}\indices{_c^\mu_a}(x)$. Similar to \eqref{eq:invariant_building:blocks_full_h}, it reads
\begin{align}
     \hat{\mathsf{E}}\indices{_c^\mu_a}(x)
     &= \lc W\indices{_\alpha^\mu} W_c^{-1}\lp \hat{E}\indices{_s^\alpha_a}(x) + \hat{E}\indices{_c^\alpha_a}(x)\rp\rc - \hat{E}\indices{_s^\mu_a}(x)\nn\\
     &=\hat{E}\indices{_s^\rho_a}(x)\bigg[-\frac{1}{2}\hat{\mathfrak{h}}\indices{_\rho^\mu}(x)+\frac{3}{8}\hat{\mathfrak{h}}\indices{_\rho_\sigma}(x)\hat{\mathfrak{h}}\indices{^\sigma^\mu}(x)\bigg]+\mathcal{O}(\hat{\mathfrak{h}}^3)\nn\\
     &= -\frac{1}{2}\hat{\mathfrak{h}}\indices{_a^\mu}(x) + \frac{3}{8}\hat{\mathfrak{h}}\indices{_a_\sigma}(x)\hat{\mathfrak{h}}\indices{^\sigma^\mu}(x) + \mathcal{O}(\hat{\mathfrak{h}}^3)\,.
 \end{align}
 In the second line, we fixed symmetric gauge for $\hat{E}\indices{_c^\alpha_a}(x)$. Since this object is then contracted with the non-symmetric $\hat{E}\indices{_s^\rho_a}(x)$, the resulting $\hat{\mathsf{E}}\indices{_c^\mu_a}(x)$ is also not symmetric.
 In addition, we defined in the last line the graviton with LLT index as
 \begin{equation}
     \hat{\mathfrak{h}}\indices{_a^\mu}(x) \equiv \hat{E}\indices{_s^\rho_a}(x)\hat{\mathfrak{h}}\indices{_\rho^\mu}(x)\,,
 \end{equation}
 as a convenient abbreviation. Since $\hat{E}\indices{_s^\rho_a}(x)$ is not symmetric, $\hat{\mathfrak{h}}\indices{_a^\mu}(x)$ is not symmetric and the $a$ must be treated as a LLT index, while the index $\mu$ is a GCT index for the purpose of raising and lowering.

The spin-connection $\hat{\Omega}_\mu(x)$ inside the covariant derivative \eqref{eq:ConstructionInvCovDev} is also constructed using this gauge-covariant vierbein
$\hat{E}\indices{_s^\mu_a}(x) + \hat{\mathsf{E}}\indices{_c^\rho_a}(x)$.
The other combination, $\mathfrak{E}\indices{_s^\sigma_b}(x)$, is the inverse vierbein corresponding to $\mathfrak{e}\indices{_s^a_\mu}(x)$ given in \eqref{eq:GaugeCovariantVierbein}, which contains the Riemann tensor terms.
It reads
\begin{equation}\label{eq:FrakVierbein}
    \mathfrak{E}\indices{_s^\mu_ a}(x) = \frac{1}{4}n \indices{_+_ a} y^\rho y^\sigma  R\indices{^ \mu_\rho_-_\sigma}+\frac{1}{6}y^\rho y^\sigma
    R \indices{^ \mu_\rho_{ a_\bot}_\sigma}+\frac{1}{12} n\indices{_-_ a}y^\rho y^\sigma 
    R \indices{^ \mu_\rho_+_\sigma} + \mathcal{O}(\lambda^5)\,.
\end{equation}

One can now pull out all collinear Wilson lines using the same tricks as for the soft ones presented in \eqref{eq:RewritingTricks} and following.
Then, adopting the discussion around (E.15) in App.\ E in \cite{Beneke:2021aip}, one can cancel the collinear GCT and LLT Wilson lines using similar gauge-invariance arguments.
The resulting Lagrangian reads\footnote{Note that in this formulation, indices are raised/lowered or changed from GCT to LLT using the emergent background metric tensor and vierbein as 
\begin{equation*}
    \hat{{h}}^{\alpha\beta}(x)=\hat{g}_s^{\alpha\mu}(x)\hat{g}_s^{\beta\nu}(x)\hat{{h}}\indices{_\mu_\nu}(x)\,,\quad
    \hat{{ h}}\indices{_a_\mu}(x)=\hat{E}\indices{_s^\nu_a}(x) \hat{{h}}\indices{_\nu_\mu}(x)\,.
\end{equation*}
Derivatives in square brackets only act within these brackets.}
\begin{align}
 \label{eq:Lagrangian_with_redefined_fields}
\mathcal{L} &= \sqrt{-\det\left(\hat{g}\indices{_s_\mu_\nu}(x)+\hat{h}\indices{_\mu_\nu}(x)\right)} \nonumber \\ 
    &\quad\times \left(\overline{\Hat{\xi}}(x)+\overline{\Hat{\zeta}}(x)+\left[W_c\left[ R^{-1}\overline{q}(x)\right]V_sV_c^{-1}\right]   \right)\gamma^a\left(\Hat{E}\indices{^\rho_a}(x)+\left[W_c W\indices{_c_\sigma^\rho}V\indices{_c_a^b}\mathfrak{E}\indices{_s^\sigma_b}(x) \right]\right)\nonumber \\
    &\quad\times i\hat{D}_\rho(x)\left( \hat{\xi}(x)+\hat{\zeta}(x)+\Bigl[W_c V_c V_s^{-1}\left[ R^{-1}q(x)\right]\Bigr]\right) +\mathcal{O}\left(\lambda^7\right)\,,
\end{align}
and is now expressed purely in terms of the hatted fields $\hat{\zeta}$, $\hat{\xi}$, $\hat{h}\indices{_\mu_\nu}$, and the soft background $\hat{e}\indices{_s^a_\mu}(x)$ and its inverse, as well as the spin-connection $\hat{\Omega}_\mu$, which now involves $\Hat{E}\indices{^\rho_a}(x)=\hat{E}\indices{_s^\rho_a}(x)+\hat{E}\indices{_c^\rho_a}(x)$, and $\hat{E}\indices{_c^\rho_a}$ is the inverse of $\hat{e}\indices{_c^a_\mu}$ defined by \eqref{eq:redefinition_fields}.
Next one can integrate out the subleading component $\hat\zeta$ using its equation of motion. 
It is again instructive to first derive this equation of motion for the case where the soft quark is set to 0, $q(x)=0$ and without the additional Riemann tensor terms.
The effect of these terms can be included in a straightforward fashion, but they lead to quite cumbersome expressions.
The final result is extended to the full expression including all relevant terms in the end. 
Omitting the argument $x$ for brevity, the equation of motion reads
\begin{equation}
    \frac{\slashed{n}_-\slashed{n}_+}{4} \gamma^a \hat{E} \indices{^\mu_a}\left( i\partial_\mu+\hat{\Omega}_{\mu}\right) \left(\hat\xi +\hat\zeta \right)=0\, .
\end{equation}
Expanding this expression using the light-cone vectors $n_\pm^\mu$, one arrives at
\begin{equation}
     \frac{\slashed{n}_-\slashed{n}_+}{4} \left( \frac{\hat{E}\indices{^\mu_+}\slashed n_-i\partial_\mu }{2} +\hat{E}\indices{^\mu_a}\gamma_\bot^a i\partial_\mu +\hat{E}\indices{^\mu_a}\gamma^a\hat{\Omega}_{\mu}\right)\left(\hat\xi +\hat\zeta \right)=0\, .
\end{equation}
Using the anticommutation relations $\left\{\slashed{n}_\pm,\gamma_\bot^a\right\}=0$, as well as the projection properties \eqref{eq:SpinorSplit}
\begin{equation}
    \hat\xi = \frac{\slashed{n}_-\slashed{n}_+}{4} \hat\xi\,,\quad \hat\zeta = \frac{\slashed{n}_+\slashed{n}_-}{4}\hat\zeta\,,
\end{equation}
which imply $\slashed{n}_-\hat{\xi}=\slashed{n}_+\hat{\zeta}=0$,
one can solve for $\hat{\zeta}$
\begin{equation}\label{eq:eom_result_zeta_coll}
    \hat{\zeta} = -\frac{\slashed{n}_+}{2}\frac{1}{\hat{E}\indices{^\mu_+}i\partial_\mu + \hat{E}\indices{^\mu_a}\hat{\omega}_{\mu bc}B^{abc}}
    \hat{E}\indices{^\mu_a}\lp \eta_\perp^{ad}i\partial_\mu + A^{abcd}\hat{\omega}_{\mu bc}\rp \gamma_{\perp d} \hat\xi\,,
\end{equation}
where we defined\footnote{These objects are the remaining structures of $\gamma^a\Sigma^{bc}$ once all projection properties are taken into account.}
\begin{align}
    A^{abcd} &= \eta_\perp^{ad}\np^{b} \nm^{c} +
    2 \np^a \nm^b \eta_\perp^{cd}
    + 2 \eta_\perp^{ab}\eta_\perp^{cd}\,,\\
    B^{abc} &= \np^a \nm^{b}\np^{c} + 2 \gamma_\perp^a \gamma_\perp^{b}\np^{c}
    + \frac 12 \np^a \lc\gamma_\perp^b,\gamma_\perp^c\rc
    \,.
\end{align}
The inverse operator is defined through its $\lambda$-expansion as
\begin{equation}\label{eq:InverseEDerivative}
    \frac{1}{\hat{E}\indices{^\mu_+}i\partial_\mu + \hat{E}\indices{^\mu_a} \hat{\omega}_{\mu bc}B^{abc}}=\frac{1}{i\partial_+}-\frac{1}{i\partial_+}\left(\bigl(\hat{E}\indices{^\mu_+}-\delta\indices{^\mu_+}\bigr)\partial_\mu+\hat{E}\indices{^\mu_a}\hat{\omega}_{\mu bc}B^{abc}\right)\frac{1}{i\partial_+}+\mathcal{O}(\lambda^2)\,,
\end{equation}
and the inverse derivative is defined as \cite{Beneke:2002ni}
\begin{equation}
\frac{1}{i\partial_+}f(x^\mu) \equiv \frac{1}{i\np\partial+i\epsilon}f(x^\mu)  = -i 
\int_{-\infty}^0 \!ds \,f(x^\mu+s \np^\mu)\,.
\end{equation}

When reinstating the soft quark-field one has to take into account momentum conservation, which implies that terms containing only a single collinear field vanish when working in GCT and LLT light-cone gauge \cite{Beneke:2002ni}.
For the case at hand, this leads to the following expression for $\hat{\zeta}$
\begin{align}
\label{eq:eta_with_soft_quark}
     \hat{\zeta}&=-\frac{\slashed n_+}{2}\left (\frac{1}{\Hat{E}\indices{^\sigma_+}i\partial_\sigma+\Hat{E}\indices{^\mu_a} \hat{\omega}_{\mu bc} B^{abc}} \right)\gamma^a \left(\hat{E}\indices{^\mu_a}i\hat{D}_\mu\right)\left[\Hat{\xi}+W_c V_c V_s^{-1}\left[ R^{-1}q(x)\right]\right]\\
     &\quad+\frac{\slashed n_+}{2}\left (\frac{1}{\Hat{E}\indices{^\sigma_+}i\partial_\sigma+\Hat{E}\indices{^\mu_a} \hat{\omega}_{\mu bc}B^{abc}} \right) \lp 1 - \frac 12 \hat{h}\indices{^\alpha_\alpha} \rp\gamma^a i\partial_a\Bigl[W_c V_c V_s^{-1}\left[ R^{-1}q(x)\right]\Bigr]+\mathcal{O}(\lambda^5)\, ,\nn
\end{align}
where the inverse derivative operator should be expanded according to \eqref{eq:InverseEDerivative}.
Above, we omitted the argument $x$ for the collinear spinor field and the vierbein to allow for a shorter notation.
Since the soft quark $q$ has no projection properties, we keep the full $\gamma^a i\hat{D}_\mu$ for conciseness in the numerator.
When acting on the $\hat\xi$ field, the structure reduces to $A^{abcd}$ as in \eqref{eq:eom_result_zeta_coll}.

Inserting the expression \eqref{eq:eta_with_soft_quark} into the Lagrangian \eqref{eq:Lagrangian_with_redefined_fields} results in
\begin{equation}
    \mathcal{L}=\sqrt{-\hat{g}_s}(x)\left(\mathcal{L}^{(0)}+\mathcal{L}^{(1)}+\mathcal{L}^{(2)} \right)\,,
\end{equation}
where the superscript denotes the leading $\lambda$-counting of each term. It is convenient to express the Lagrangian in terms of the GCT gauge-invariant building blocks $\hat{\chi}_c=W_c^{-1}\hat{\xi}$  and $\hat{\mathfrak{h}}_{\mu\nu}$.
At leading and subleading order, the terms read
\begin{align}
    \mathcal{L}^{(0)} &=\overline{\hat{\chi}}_c\frac{\slashed{n}_+}{2}\lp i\nm D_s +i\slashed{\partial}_\bot \frac{1}{i \np\partial}i\slashed{\partial}_\bot\rp \hat{\chi}_c\,, \label{eq:L_0}\\
    \mathcal{L}^{(1)}&=
    -\frac{1}{4}\overline{\hat{\chi}}_c\frac{\slashed{n}_+}{2}
    \lp
    \left\{\nm^a\hat{\mathfrak{h}}\indices{_a^\mu}\,,i\partial_\mu \right\}+
    \gamma_\perp^a\left\{\hat{\mathfrak{h}}\indices{_a^\rho}\,,i\partial_\rho\right\}\frac{1}{i\np\partial}i\slashed{\partial}_\bot + i\slashed{\partial}_\bot \frac{1}{i\np\partial}\gamma_\perp^a\left\{\hat{\mathfrak{h}}\indices{_a^\rho}\,,i\partial_\rho\right\}\rp \hat{\chi}_c\nn\\
    &\quad+\frac{1}{4}\hat{\mathfrak{h}}\left(
    \overline{\hat{\chi}}_c\frac{\slashed{n}_+}{2}\Bigl(i \nm D_s
    + i\slashed{\partial}_\bot \frac{1}{i\np\partial}i\slashed{\partial}_\bot \Bigr)\hat{\chi}_c + \mathrm{h.c.}
    \right)\,.
\end{align}
Here, $\{\cdot,\cdot\}$ denotes the anticommutator and $\hat{\mathfrak{h}}\equiv\hat{\mathfrak{h}}\indices{^\alpha_\alpha}$.
At the second order, one has
\begin{align}
\label{eq:Lagrangian_order_2}
    \mathcal{L}^{(2)}&=\frac{1}{16}\overline{\hat{\chi}}_c\frac{\slashed n_+}{2}
    \lp
    \Bigl\{ {\hat{\mathfrak{h}}^2 
       - 2 \hat{\mathfrak{h}}^{\alpha\beta}\hat{\mathfrak{h}}_{\alpha\beta}}\,,i\nm D_s\Bigr\}
    +\Bigl\{ {\nm^a\bigl(3 \hat{\mathfrak{h}}^{\mu\alpha}\hat{\mathfrak{h}}_{\alpha a}- 2 \hat{\mathfrak{h}} \hat{\mathfrak{h}}\indices{^\mu_a}\bigr)}\,,i\partial_\mu\Bigr\}\rp\hat{\chi}_c\nonumber\\
        &\quad-\overline{\hat{\chi}}_c\frac{\slashed n_+}{2}\Biggl(
       \frac{1}{2}\Bigl\{i\slashed\partial_\bot\frac{1}{in_+\partial}\,, \gamma_\bot^a \bigl\{{\mathcal G\indices{^\mu_a}}\,,i\partial_\mu\bigr\} 
       \Bigr\}\nonumber\\
       &\quad\phantom{-\overline{\hat{\chi}}_c\frac{\slashed n_+}{2}\Biggl(}
       +\frac{1}{4}\Bigl(i\slashed\partial_\bot \Bigl\{\frac{1}{in_+\partial}\,,{\hat{\mathfrak{h}}}\Bigr\} \gamma^a_\bot \bigl\{{\mathcal H\indices{^\alpha_a}}\,, i\partial_\alpha\bigr\}
       + \gamma^a_\bot \bigl\{{\mathcal H\indices{^\alpha_a}}\,, i\partial_\alpha\bigr\}\Bigl\{\frac{1}{in_+\partial}\,,{\hat{\mathfrak{h}}} \Bigr\}i\slashed\partial_\bot\Bigr)\nonumber\\
       &\quad\phantom{-\overline{\hat{\chi}}_c\frac{\slashed n_+}{2}\Biggl(}
       - \frac{1}{4}\gamma_\bot^a \bigl\{ {\mathcal H\indices{^\mu_a}}\,, i\partial_\mu\bigr\}\frac{1}{in_+\partial} \bigl\{{\mathcal H\indices{^\alpha_b}}\,,i\partial_\alpha\bigr\} \gamma_\bot^b
       + \frac{1}{16}i\slashed\partial_\bot \Bigl\{\frac{1}{in_+\partial}\,, {\hat{\mathfrak{h}}}^2 - 2 {\hat{\mathfrak{h}}^{\alpha\beta}\hat{\mathfrak{h}}_{\alpha\beta}}\Bigr\}i\slashed\partial_\bot\nonumber\\
       &\quad\phantom{-\overline{\hat{\chi}}_c\frac{\slashed n_+}{2}\Biggl(}
       - \frac{1}{16}i\slashed\partial_\bot \biggl(\Bigl\{\frac{1}{in_+\partial}\,, {\hat{\mathfrak{h}}}^2\Bigr\} + \frac{3}{2}{\hat{\mathfrak{h}}}^2\frac{1}{in_+\partial} + {\hat{\mathfrak{h}}}\Bigl\{\frac{1}{in_+\partial}\,, {\hat{\mathfrak{h}}}\Bigr\}\biggr)i\slashed\partial_\bot\nonumber\\
       &\quad\phantom{-\overline{\hat{\chi}}_c\frac{\slashed n_+}{2}\Biggl(}
       -\frac{1}{32}\biggl( {\hat{\mathfrak{h}}\indices{^\lambda_a}} [i\partial_b {\hat{\mathfrak{h}}_{c\lambda}}]\bigl(\nm^c\lc\gamma_\perp^a,\gamma_\perp^b\rc + \nm^b\lc\gamma_\perp^c,\gamma_\perp^a\rc + \nm^a\lc\gamma_\perp^b\,,\gamma_\perp^c\rc\bigr)\nn\\
       &\quad\phantom{-\overline{\hat{\chi}}_c\frac{\slashed n_+}{2}\Biggl(}
       \phantom{-\frac{1}{32}\biggl(}
       +\Bigl[i\slashed\partial_\bot\frac{1}{in_+\partial}\,, {\hat{\mathfrak{h}}\indices{^\lambda_a}} [i\np\partial {\hat{\mathfrak{h}}_{c\lambda}}]\bigl(\gamma_\perp^c \nm^a - \gamma_\perp^a \nm^c\bigr)\Bigr]\nn\\
       &\quad\phantom{-\overline{\hat{\chi}}_c\frac{\slashed n_+}{2}\Biggl(}
       \phantom{-\frac{1}{32}\biggl(}
       +i\slashed\partial_\bot \frac{1}{in_+\partial} {\hat{\mathfrak{h}}\indices{^\lambda_a}} [i\np\partial {\hat{\mathfrak{h}}_{c\lambda}}]\lc\gamma_\perp^a\,,\gamma_\perp^c\rc\frac{1}{in_+\partial}i\slashed\partial_\bot\biggr)\Biggr){\hat{\chi}_c}\nn\\
       &\quad
       +\frac{1}{4}x_\perp^\alpha x_\perp^\beta R\indices{_-_\alpha_-_\beta}\,\overline{\hat{\chi}}_c\frac{\slashed{n}_+}{2} i\np\partial \hat{\chi}_c\nn\\
       &\quad-\frac{1}{4}\lp\overline{q}\lp\left\{\hat{\mathfrak{h}}\indices{^\mu_a}\,,i\partial_\mu\right\}\gamma_\perp^a
       - i\slashed{\partial}_\bot\frac{1}{i{\np\partial}}\left[i\np\partial\hat{\mathfrak{h}}\right]\rp\hat{\chi}_c+ \mathrm{h.c.}\rp\,,
\end{align}
where 
\begin{equation}
\mathcal{H}\indices{^\mu_\nu}=\frac{1}{2} \left( \hat{\mathfrak{h}}\indices{^\mu_\nu}-\hat{\mathfrak{h}} \delta \indices{^\mu_\nu}\right)\,,\quad
\mathcal{G}\indices{^\mu_\nu }= 
       -\frac{1}{8}\Bigl( 3 \hat{\mathfrak{h}}^{\mu\alpha}\hat{\mathfrak{h}}_{\alpha\nu} 
       - 2 \hat{\mathfrak{h}} \hat{\mathfrak{h}}\indices{^\mu_\nu} 
       + \bigl(\hat{\mathfrak{h}}^2 
       - 2 \hat{\mathfrak{h}}^{\alpha\beta}\hat{\mathfrak{h}}_{\alpha\beta}\bigr)\delta\indices{^\mu_\nu}\Bigr)\,.
\end{equation}
The Riemann tensor term in the second-to-last line in $\mathcal{L}^{(2)}$ originates from the expression
\begin{equation}
    \mathcal{L}\supset \overline{\hat{\chi}}_c [V_c]\indices{^a_b}\mathfrak{E}\indices{_s^\mu_a}\gamma^b i \hat{D}_\mu\hat{\chi}_c\,,
\end{equation}
in \eqref{eq:Lagrangian_with_redefined_fields} and we used that $[V_c]\indices{^a_b} = \delta\indices{^a_b} + \mathcal{O}(\lambda)$ to simplify the expressions.
At higher orders, there will be contributions from the collinear LLT Wilson line to these gauge-covariant Riemann tensor terms, similar to the $W_c$ Wilson line that appears with the soft field-strength tensor in QCD.
The Lagrangian can be drastically simplified using the equations of motion of both the collinear matter field and the collinear graviton.
    The trace $\hat{\mathfrak{h}}$ of the collinear graviton vanishes at $\mathcal{O}(\lambda)$, and enters at $\mathcal{O}(\lambda^2)$ via the equation \cite{Beneke:2022pue}
    \begin{align}
        \hat{\mathfrak{h}} &= \frac 12 \biggl(\hat{\mathfrak{h}}_{\alpha\beta}\hat{\mathfrak{h}}^{\alpha\beta}- \frac{1}{(\np\partial)^2}\Bigl([\np\partial\hat{\mathfrak{h}}_{\alpha\beta}]\np\partial\hat{\mathfrak{h}}^{\alpha\beta} - \overline{\hat{\chi}}_c\frac{\slashed{n}_+}{2} i\np\overset{\leftrightarrow}{\partial}\hat{\chi}_c
        \Bigr)\biggr)+\mathcal{O}(\lambda^3)\,,
    \end{align}
    where $\overline{\hat{\chi}}_c\frac{\slashed{n}_+}{2} i\np\overset{\leftrightarrow}{\partial}\hat{\chi}_c \equiv \overline{\hat{\chi}}_c\frac{\slashed{n}_+}{2} i\np\partial\hat{\chi}_c - \lc i\np\partial \overline{\hat{\chi}}_c\rc\frac{\slashed{n}_+}{2}\hat{\chi}_c$.
    This means that the terms containing the trace $\hat{\mathfrak{h}}$ in $\mathcal{L}^{(2)}$ can be dropped.
    In $\mathcal{L}^{(1)}$, this procedure generates new terms that would contribute to $\mathcal{L}^{(2)}$.
    However, note that the terms in $\mathcal{L}^{(1)}$ that contain the trace originate from the expansion of the determinant $\sqrt{-g}$ and take the form $\hat{\mathfrak{h}}\mathcal{L}^{(0)}$,
    proportional to the leading-order Lagrangian.
    By using the leading-power collinear matter equations of motion, this term can be seen to be suppressed by yet another power in $\lambda$, and thus these terms only contribute to $\mathcal{L}^{(3)}$.
    The same holds for the terms proportional to $(\hat{\mathfrak{h}}^2 - 2\hat{\mathfrak{h}}^{\alpha\beta}\hat{\mathfrak{h}}_{\alpha\beta})\mathcal{L}^{(0)}$ in $\mathcal{L}^{(2)}.$
    
    The simplified Lagrangian then reads
    \begin{align}
    \mathcal{L}^{(0)} &=\overline{\hat{\chi}}_c \,\frac{\slashed{n}_+}{2}\biggl( i\nm D_{s} \hat{\chi} +i\slashed{\partial}_\bot \frac{1}{i\np\partial}i\slashed{\partial}_\bot\biggr)\hat{\chi}_c\,, \label{eq:newL_0}\\
    \mathcal{L}^{(1)}&=-\frac{1}{4}\overline{\hat{\chi}}_c\,\frac{\slashed{n}_+}{2}\biggl(
    \left\{\nm^a\hat{\mathfrak{h}}\indices{_a^\mu},i\partial_\mu \right\}+ \gamma_\perp^a\left\{\hat{\mathfrak{h}}\indices{_a^\rho},i\partial_\rho\right\}\frac{1}{i\np\partial}i\slashed{\partial}_\bot+i\slashed{\partial}_\bot \frac{1}{i\np\partial}\gamma_\perp^a\left\{\hat{\mathfrak{h}}\indices{_a^\rho},i\partial_\rho\right\}\biggr)\hat{\chi}_c\, ,
\end{align}
which reproduces the result of \cite{Beneke:2012xa}.
At the second order, one has
\begin{align}
\label{eq:newLagrangian_order_2}
    \mathcal{L}^{(2)}&=
    \overline{\hat{\chi}}_c\,\frac{\slashed{n}_+}{2} \Biggl( 
    \frac{3}{16}\bigl\{n_-^a \hat{\mathfrak{h}}_{a\alpha}\hat{\mathfrak{h}}^{\alpha\mu}\,,i\partial_\mu\bigr\}
    +\frac{3}{16}\Bigl\{i\slashed\partial_\perp\frac{1}{i\np\partial}\,,\bigl\{\gamma_\perp^a \hat{\mathfrak{h}}_{a\alpha}\hat{\mathfrak{h}}^{\alpha\mu}\,,i\partial_\mu\bigr\}\Bigr\}\nn\\
    &\quad\phantom{\overline{\hat{\chi}}_c\,\frac{\slashed{n}_+}{2} \Biggl(}
    +\frac{1}{16}\gamma_\perp^a \bigl\{\hat{\mathfrak{h}}\indices{_a^\mu}\,,i\partial_\mu\bigr\}\frac{1}{i\np\partial}\gamma_\perp^b\bigl\{\hat{\mathfrak{h}}\indices{_b^\nu}\,,i\partial_\nu\bigr\}\nn\\
    &\quad\phantom{\overline{\hat{\chi}}_c\,\frac{\slashed{n}_+}{2} \Biggl(}
        -\frac{1}{32}\biggl( {\hat{\mathfrak{h}}\indices{^\lambda_a}} [i\partial_b {\hat{\mathfrak{h}}_{c\lambda}}](\nm^c\lc\gamma_\perp^a,\gamma_\perp^b\rc + \nm^b\lc\gamma_\perp^c\,,\gamma_\perp^a\rc + \nm^a\lc\gamma_\perp^b,\gamma_\perp^c\rc)\nn\\
        &\quad\quad\phantom{\overline{\hat{\chi}}_c\,\frac{\slashed{n}_+}{2} \Biggl(}
        \phantom{-\frac{1}{32}\biggl(}
       +\Bigl[i\slashed\partial_\bot\frac{1}{in_+\partial}\,, \hat{\mathfrak{h}}\indices{^\lambda_a} [i\np\partial {\hat{\mathfrak{h}}_{c\lambda}}](\gamma_\perp^c \nm^a - \gamma_\perp^a \nm^c)\Bigr]\nn\\
       &\quad\quad\phantom{\overline{\hat{\chi}}_c\,\frac{\slashed{n}_+}{2} \Biggl(}
        \phantom{-\frac{1}{32}\biggl(}
       +i\slashed\partial_\bot \frac{1}{in_+\partial} {\hat{\mathfrak{h}}\indices{^\lambda_a}} [i\np\partial {\hat{\mathfrak{h}}_{c\lambda}}]\lc\gamma_\perp^a\,,\gamma_\perp^c\rc\frac{1}{in_+\partial}i\slashed\partial_\bot\biggr)\Biggr){\hat{\chi}_c}\nn\\
       &\quad+\frac{1}{4}x_\perp^\alpha x_\perp^\beta R\indices{_-_\alpha_-_\beta}\overline{\hat{\chi}}_c\frac{\slashed{n}_+}{2} i\np\partial \hat{\chi}_c
        -\lp\frac{1}{4}\overline{q}\left\{\hat{\mathfrak{h}}\indices{^\mu_a}\,,i\partial_\mu\right\}\gamma_\perp^a \hat{\chi}_c + \mathrm{h.c.}\rp\,.
\end{align}
In this form, one notices that the soft-covariant derivative $\nm D_s$ only appears in the leading-order Lagrangian $\mathcal{L}^{(0)}$ due to the light-cone gauge condition $\hat{\mathfrak{h}}_{\mu+}=0$.

Eqs.~\eqref{eq:newL_0} -- \eqref{eq:newLagrangian_order_2} 
are the main results of this work. They demonstrate that 
the developed formalism allows one to compute in a systematic way soft and collinear graviton interactions to any order 
in the soft and collinear limits for matter fields with spin. There are no collinear 
interactions at leading order in this expansion, which reflects the absence of collinear singularities in graviton emission. When comparing the above result to the 
corresponding one in non-abelian gauge theory \cite{Beneke:2002ni}, one further notes that there are no interactions with 
the soft matter field and the Riemann tensor at 
$\mathcal{O}(\lambda)$ contrary to the case of gauge fields---these interactions appear only at 
$\mathcal{O}(\lambda^2)$ in gravity. The only subleading power 
$\mathcal{O}(\lambda)$ appear in the covariant derivative 
$\nm D_s$ in the  $\mathcal{L}^{(0)}$ term, which contains a 
subleading eikonal coupling to the background 
spin-connection. 

\section{Conclusion}

In this work we extended the construction of SCET gravity to include half-integer spin matter fields.
We introduced the necessary Wilson lines and redefinitions that define collinear fields with a ``homogeneous'' gauge transformation under local Lorentz transformations.
In combination with the previous results for the scalar field, which directly apply also to the diffeomorphism-transformation of half-integer spin fields, one obtains all the necessary ingredients to perform the systematic construction of the subleading Lagrangian to arbitrary order in $\lambda$ for fields of both integer and half-integer spin. The basic structure of soft-collinear gravity, which features a homogeneous soft background field, giving rise to a covariant derivative and multipole-expanded covariant Riemann-tensor interactions, remains unaltered and generalises in a natural way to fermion fields. In particular, the coupling of the subleading background gauge field to the angular momentum operator  is seen to consistently include the spin operator.\\[0.2cm]

\subsubsection*{Acknowledgement}
This work was supported by the Excellence Cluster ORIGINS
funded by the Deutsche Forschungsgemeinschaft 
(DFG)
under Grant No. EXC - 2094 - 390783311
and
the Cluster of Excellence PRISMA\textsuperscript{+} funded by the DFG 
under Grant No. EXC - 2118 - 390831469. 
MB thanks the Albert Einstein Center at the University of 
Bern for hospitality when this work was completed.
    
\bibliography{References.bib}

\end{document}